\documentclass[journal]{IEEEtran}

\IEEEoverridecommandlockouts

\makeatletter

\def\ie{\textit{i.e.}}
\def\etal{\textit{et al.}}

\def\eg{\textit{e.g.}}

\makeatletter
\let\NAT@parse\undefined
\makeatother

\usepackage{cite}
\usepackage{url}
\usepackage{soul}
\usepackage{graphicx}
\usepackage{subfigure}
\usepackage{amsmath}
\usepackage{amssymb}
\usepackage{pifont}
\usepackage{array}
\usepackage{multirow}
\usepackage{color,colortbl}
\usepackage{tikz}

\usepackage{hyperref}
\hypersetup{colorlinks, citecolor=green, filecolor=pink, linkcolor=red, urlcolor=blue }

\newcommand*\circled[1]{\tikz[baseline=(char.base)]{
            \node[shape=circle,draw,inner sep=0.5pt] (char) {#1};}}

\begin{document}
\bibliographystyle{IEEEtran}

\title{Blockchain for Internet of Things: A Survey}

\author{
Hong-Ning Dai,~\IEEEmembership{Senior Member,~IEEE},~Zibin~Zheng,~\IEEEmembership{Senior Member,~IEEE},~Yan Zhang,~\IEEEmembership{Senior Member,~IEEE}
\thanks{\it Corresponding authors: Zibin Zheng and Yan Zhang.}
\thanks{H.-N. Dai is with Faculty of Information Technology, Macau University of Science and Technology, Macau (email: hndai@ieee.org).}
\thanks{Z. Zheng is with School of Data and Computer Science, Sun Yat-sen University, China (email: zhzibin@mail.sysu.edu.cn).}
\thanks{Y. Zhang is with Department of Informatics, University of Oslo, Norway. He is also with Simula Metropolitan Center for Digital Engineering, Norway (email: yanzhang@ieee.org).}
}

\maketitle


\begin{abstract}

Internet of Things (IoT) is reshaping the incumbent industry to smart industry featured with data-driven decision-making. However, intrinsic features of IoT result in a number of challenges such as decentralization, poor interoperability, privacy and security vulnerabilities. Blockchain technology brings the opportunities in addressing the challenges of IoT. In this paper, we investigate the integration of blockchain technology with IoT. We name such synthesis of blockchain and IoT as Blockchain of Things (BCoT). This paper presents an in-depth survey of BCoT and discusses the insights of this new paradigm. In particular, we first briefly introduce IoT and discuss the challenges of IoT. Then we give an overview of blockchain technology. We next concentrate on introducing the convergence of blockchain and IoT and presenting the proposal of BCoT architecture. We further discuss the issues about using blockchain for 5G beyond in IoT as well as industrial applications of BCoT. Finally, we outline the open research directions in this promising area.

\end{abstract}


\section{Introduction} 
\label{sub:intro}

The recent advances in information and communication technology (ICT) have promoted the evolution of conventional computer-aided industry to \emph{smart industry} featured with data-driven decision making \cite{Lade:IEEEIS2018}. During this paradigm shift, Internet of Things (IoT) plays an important role of connecting the physical industrial environment to the cyberspace of computing systems consequently forming a Cyber-Physical System (CPS). IoT can support a wide diversity of industrial applications such as manufacturing, logistics, food industry and utilities. IoT aims to improve operation efficiency and production throughput, reduce the machine downtime and enhance product quality. In particular, IoT has the following features: 1) decentralization of IoT systems, 2) diversity of IoT devices and systems, 3) heterogeneity of IoT data and 4) network complexity. All of them result in the challenges including heterogeneity of IoT system, poor interoperability, resource constraints of IoT devices, privacy and security vulnerabilities.

The appearance of blockchain technologies brings the opportunities in overcoming the above challenges of IoT. A blockchain is essentially a distributed \emph{ledger} spreading over the whole distributed system. With the decentralized consensus, blockchains can enable a transaction to occur and be validated in a mutually-distrusted distributed system without the intervention of the trusted third party. Unlike incumbent transaction-management systems where the  centralized agency needs to validate the transaction, blockchains can achieve the \emph{decentralized} validation of transactions, thereby greatly saving the cost and mitigating the performance bottleneck at the central agency. Moreover, each transaction saved in blockchains is essentially \emph{immutable} since each node in the network keeps all the committed transactions in the blockchain. Meanwhile, crytographic mechanisms (such as asymmetric encryption algorithms, digital signature and hash functions) guarantee the integrity of data blocks in the blockchains. Therefore, the blockchains can ensure non-repudiation of transactions. In addition, each transaction in blockchains is traceable to every user with the attached historic timestamp. 

Blockchain is essentially a perfect complement to IoT with the improved interoperability, privacy, security, reliability and scalability. In this paper, we investigate a new paradigm of integrating blockchain with IoT. We name such synthesis of blockchain and IoT as Blockchain of Things (BCoT). In particular, BCoT has the following merits:
\begin{itemize}
\item \emph{Interoperability} across IoT devices, IoT systems and industrial sectors, where the interoperability is the ability of interacting with physical systems and exchanging information between IoT systems. It can be achieved through the \emph{blockchain-composite layer} built on top of an overlay peer-to-peer (P2P) network with uniform access across different IoT systems.

\item \emph{Traceability} of IoT data, where the traceability is the capability of tracing and verifying the spatial and temporal information of a data block saved in the blockchain. Each data block saved in a blockchain is attached with a historic timestamp consequently assuring the data traceability.

\item \emph{Reliability} of IoT data is the quality of IoT data being trustworthy. It can be ensured by the integrity enforced by crytographic mechanisms including asymmetric encryption algorithms, hash functions and digital signature, all of which are inherent in blockchains.

\item \emph{Autonomic interactions} of IoT system refer to the capability of IoT systems interacting with each other without the intervention of a trusted third party. This autonomy can be achieved by \emph{smart contracts} enabled by blockchains. In particular, contract clauses embedded in smart contracts will be executed automatically when a certain condition is satisfied (\eg, the user breaching the contract will be punished with a fine automatically).
\end{itemize}
 
Though BCoT can benefit IoT, there are also a number of challenges to be addressed before the potentials of BCoT can be fully unleashed. Therefore, this paper aims to present an in-depth survey on the state-of-the-art advances, challenges and open research issues in BCoT. 


\subsection{Comparison between this paper and existing surveys}
There are several published papers discussing the convergence of blockchain with IoT. For example, the work of \cite{Dorri:PercomWorkshp2017} presents a smart home application of using blockchains for IoT. Zhang and Wen \cite{Zhang:P2PNA2017} proposed a business model to support P2P trading based on smart contracts and blockchains. However, these studies are too specific to a certain scenario of incorporating blockchain with IoT (\eg, a smart home application).

Recently, several surveys on the convergence of blockchain with IoT have been published. In particular, \cite{Conoscenti:AICCSA16} gives a systematic literature review on blockchain for IoT with the categorization of a number of use cases. The work of \cite{BANERJEE:2017} presents a survey on IoT security and investigates the potentials of blockchain technologies as the solutions. Reyna \etal \cite{REYNA:2018} investigated the possibility and research issues of integrating blockchain with IoT. The work of \cite{Fernandez-Carames:Access2018} presents a review on integrating blockchain with IoT in the application aspect. Ref. \cite{MSAli:CST2018} attempted to give a comprehensive survey on application of blockchain in IoT. The work of \cite{Panarello:Sensors18} gives a categorization of applications of blockchain for IoT. 

However, most of the existing surveys suffer from the following limitations: 1) there is no general architecture proposed for BCoT; 2) there is no study explicitly discussing blockchain for 5G beyond networks for IoT (however, this topic is of great importance for the development of IoT); 3) other important issues like life cycle of smart contracts are missing in most of the existing surveys.

\subsection{Contributions}
In view of prior work, we aim to (i) provide a conceptual introduction on IoT and blockchain technologies, (ii) present in-depth analysis on the potentials of incorporating blockchains into IoT and (iii) give insightful discussions of technical challenges enabling BCoT. In summary, the main contributions of this paper are highlighted as follows:
\begin{enumerate}
\item A brief introduction on IoT is first given and then accompanied by a summary of key characteristics of IoT. Meanwhile, research challenges of IoT are outlined. 

\item An overview of key blockchain technologies is then given with a summary of key characteristics of blockchains and a taxonomy of the incumbent blockchain systems. 

\item The core part of this paper is focused on the convergence of blockchain and IoT. In this respect, the opportunities of integrating blockchain with IoT are first discussed. An architecture of BCoT is then proposed and illustrated. 

\item The 5G-beyond networks play an important role in constructing the infrastructure for BCoT. Research issues about blockchain for 5G-beyond networks in IoT are also discussed. 

\item Furthermore, this paper summarizes the applications of BCoT and outlines the open research issues in BCoT. 
\end{enumerate}

The remainder of the paper is organized as follows. Section \ref{sec:IIoT} first presents an overview on IoT. Section \ref{sec:blockchain} then gives the introduction of blockchain technology. The convergence of blockchain and IoT is discussed in Section \ref{sec:IBoT}. Section \ref{sec:5G} discusses the research issues about blockchain for 5G-beyond networks. Section \ref{sec:IBoTapp} next summarizes the applications of BCoT. Open research issues are discussed in Section \ref{sec:chall-ibot}. Finally, the paper is concluded in Section \ref{sec:conc}.

\section{Internet of Things} 
\label{sec:IIoT}

In this section, we briefly introduce Internet of Things (IoT) in Section \ref{subsec:intro-IIoT} and summarize the challenges of IoT in Section \ref{subsec:challenges-IIoT}.

\subsection{Introduction to Internet of Things}
\label{subsec:intro-IIoT}
Today's industry is experiencing a paradigm shift from conventional computer-aided industry to \emph{smart industry} driven by recently advances in Internet of Things (IoT) and Big Data Analytics (BDA). During this evolution, IoT plays a critical role of bridging the gap between the physical industrial environment and the cyberspace of computing systems while BDA can help to extract hidden values from massive IoT data so as to make intelligent decisions.

IoT is essentially a network of smart objects (\ie, things) with provision of various industrial services. A typical IoT system consists of the following layered sub-systems (from bottom to up) as shown in Fig. \ref{fig:IIoT}:
\begin{itemize}

\item \emph{Perception Layer}: There is a wide diversity of IoT devices including sensors, actuators, controllers, bar code/Quick Response Code (QR Code) tags, RFID tags, smart meters and other wireless/wired devices. These devices can sense and collect data from the physical environment. Meanwhile, some of them (like actuators and controllers) can make actions on the environment.

\item \emph{Communication Layer}: Various wireless/wired devices such as sensors, RFIDs, actuators, controllers and other tags can then connect with IoT gateways, WiFi Access Points (APs), small base stations (BS) and macro BS to form an industrial network. The network connection is enabled by a diverse of communication protocols such as Bluetooth, Near Field Communications (NFC), Low-power Wireless Personal Area Networks (6LoWPAN), Wireless Highway Addressable Remote Transducer (WirelessHART) \cite{Petersen:IEEE2011}, Low Power Wide Area Networks (LPWAN) technologies including Sigfox, LoRa, Narrowband IoT (NB-IoT) and industrial Ethernet \cite{MEKKI2018}.

\item \emph{Industrial Applications}: IoT can be widely used to support a number of industrial applications. The typical industrial applications include manufacturing, supply chain, food industry, smart grid, health care and internet of vehicles.

\begin{figure}[t]
\centering
\includegraphics[width=8.9cm]{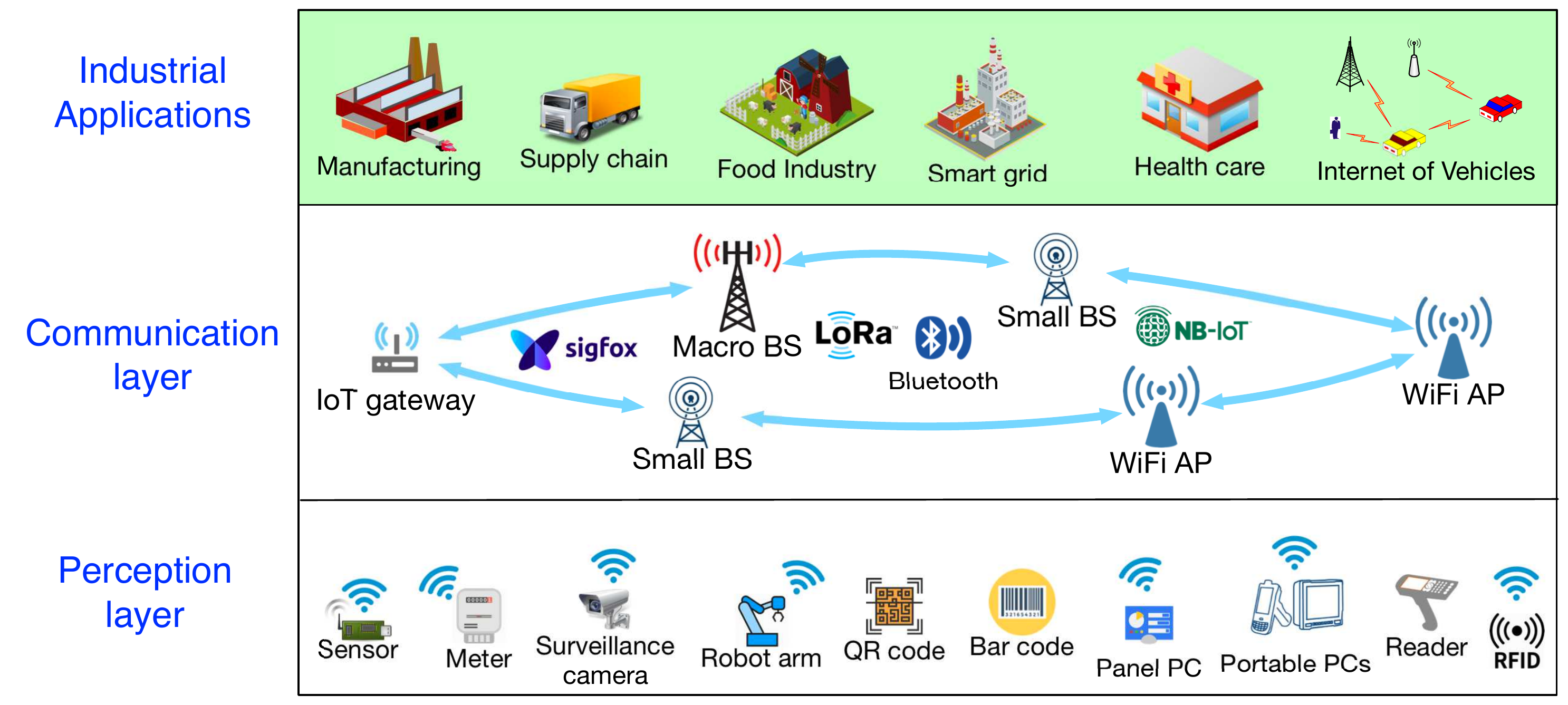}
\caption{Internet of Things (IoT) consists of perception layer, communication layer and industrial applications}
\label{fig:IIoT}
\end{figure}

\end{itemize}

\subsection{Challenges of Internet of Things}
\label{subsec:challenges-IIoT}

In this paper, we mainly focus on Industrial IoT. We denote Industrial IoT by IoT thereafter without loss of generality. The IoT ensures the connection of various \emph{things} (smart objects) mounted with various electronic or mechanic sensors, actuators and software systems which can sense and collect information from the physical environment and then make actions on the physical environment. The unique features of IoT pose a number of research challenges exhibiting in the following aspects.

\begin{figure*}[t]
\centering
\includegraphics[width=18cm]{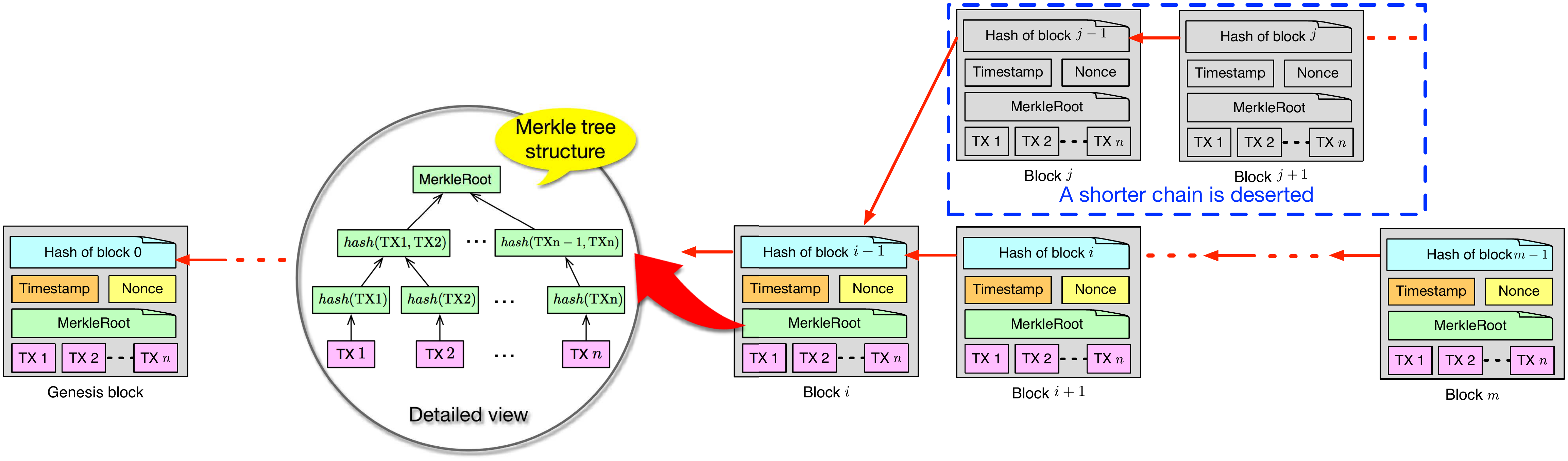}
\caption{Blockchain consists of a number of consecutively-connected blocks and the detailed view represents a Merkle tree structure (where TX represents a transaction)}
\label{fig:blockchain}
\end{figure*}

\begin{itemize}

\item \emph{Heterogeneity} of IoT systems exhibits in the heterogeneous IoT devices, heterogeneous communication protocols and heterogeneous IoT data types (\ie, structured, semi-structured and nonstructured). The heterogeneity is also the root of other challenges such as interoperability, privacy and security (to be explained as follows). 

\item \emph{Complexity} of networks. There are a number of communication/network protocols coexisting in IoT. Typical network protocols include NFC, Bluetooth, 6LoWPAN, WirelessHART, Sigfox, LoRa and NB-IoT, all of which offer different network services. For example, 6LoWPAN and WirelessHART have typically short communication coverage (\eg, less than 100 meters) while LPWAN technologies can provide the coverage from 1km to 10 km \cite{MChen:IEEEAccess2017,Khutsoane:IECON2017,hndai:EIS19}. 

\item \emph{Poor interoperability} is the capability of IoT systems (both hardware and software) to exchange, make use of information and collaborate with each other. Due to the decentralization of IoT systems and the heterogeneity of IoT systems, it is challenging to exchange the data between different industrial sectors, strategic centers, IoT systems. As a result, the interoperability of IoT is difficult to be achieved. 

\item \emph{Resource constraints of IoT devices}. IoT devices such as sensors, actuators, RFID tags and smart meters suffer from limited resources including computing resource, storage resource and battery power. For example, there is no battery power for passive RFID tags that can only harvest the energy from RFID readers or from ambient environment \cite{XLu:IEEEWC2018}. Moreover, the resource constraints also result in the vulnerability of IoT devices to malicious attacks. 

\item \emph{Privacy vulnerability}. Privacy is to guarantee the appropriate usage of IoT data while there is no disclosure of user private information without user consent. It is challenging to preserve data privacy in IoT due to the complexity and the decentralization of IoT systems, the heterogeneity of IoT systems. Moreover, it becomes a trend to integrate IoT with cloud computing since cloud computing can empower IoT with extra computing and storage capabilities. However, uploading the confidential IoT data to the third-party cloud servers may also compromise the vulnerable privacy of IoT \cite{JZhou:ComMag2017}. 

\item \emph{Security vulnerability}. The decentralization and the heterogeneity of IoT systems also result in the difficulty in ensuring the security of IoT while the security is extremely important for an enterprise. The typical solutions such as authentication, authorization and communication encryption may not be appropriate to IoT due to the difficulty in implementing the security countermeasures in resource-constrained IoT systems. Moreover, IoT systems are also vulnerable to malicious attacks due to the failure of security firmware updates in time \cite{Roman:CN2013}.

\end{itemize}

\emph{Discussion.} Some intrinsic limitations of IoT can be overcome via recent ICT advances. For example, ambient backscatter assisted communications \cite{XLu:IEEEWC2018} can help IoT nodes obtain extra energy from ambience. Meanwhile, mobile edge computing can extend the capability of IoT nodes via offloading the computationally-intensive tasks to edge servers \cite{JHe:IoTJ18}. Moreover, the recent advances in blockchain technologies offer potential solutions to the challenges such as poor interoperability, privacy and security vulnerabilities. In addition, blockchain is also beneficial to improve heterogeneity of IoT systems. We will discuss these opportunities brought by blockchain to IoT in Section \ref{subsec:opp} after giving a briefing on blockchain technologies in Section \ref{sec:blockchain}.

\section{Blockchain Technologies} 
\label{sec:blockchain}

In this section, we first give an overview on blockchain technologies in Section \ref{subsec:overview-blockchain}, then summarize the key blockchain characteristics in Section \ref{subsec:keychar} and present a taxonomy of blockchain platforms in Section \ref{subsec:taxonomy}.

\subsection{Overview of Blockchain Technologies}
\label{subsec:overview-blockchain}

\subsubsection{Blockchain}
A blockchain is essentially a distributed \emph{ledger} spreading over the whole blockchain system \cite{zibin2016blockchain}. Fig. \ref{fig:blockchain} shows an exemplary blockchain consisting of a number of consecutively-connected blocks. Each block (with the exception of the first block) in a blockchain points to its immediately-previous block (called parent block) via an \emph{inverse} reference that is essentially the \emph{hash} value of the parent block. For example, block $i$ contains the hash of block $i-1$ as shown in Fig. \ref{fig:blockchain}. The first block of a blockchain is called the \emph{genesis} block having no parent block. In particular, a block structure consists of the following information: 1) block version (indicating the validation rules to follow), 2) the hash of parent block, 3) Timestamp recording the current time in seconds, 4) Nonce staring from 0 and increasing for every hash calculation, 5) the number of transactions, 6) MerkleRoot (\ie, the hash value of the root of a Merkel tree with concatenating the hash values of all the transactions in the block) as shown in the detailed view of Fig. \ref{fig:blockchain}. 

A blockchain is continuously growing with the transactions being executed. When a new block is generated, all the nodes in the network will participate in the block validation. A validated block will be automatically appended at the end of the blockchain via the inverse reference pointing to the parent block. In this manner, any unauthorized alterations on the previously-generated block can be easily detected since the hash value of the tampered block is significantly different from that of the unchanged block. Moreover, since the blockchain is distributed throughout the whole network, the tampering behavior can also be easily detected by other nodes in the network.

\emph{Data integrity guarantee in blockchain.}
Blockchains leverage cryptographic techniques to guarantee data integrity. In particular, there are two mechanisms in blockchains to ensure the data integrity: 1) \emph{an ordered link list structure of blocks}, in which each newly-appended block must include the hash value of the preceding block. In this manner, a falsification on any of the previous blocks will invalidate the subsequent blocks. 2) \emph{Merkel Tree structure}, in which each block contains a root hash of a Merkel tree of all the transactions. Each non-leave node is essentially a hash value of two concatenated values of its two children. Therefore, a Merkel tree is typically a binary tree. In this way, any falsification on the transactions will lead to a new hash value in the above layer, consequently resulting in a falsified root hash. As a result, any falsification can be easily detected.

\subsubsection{Consensus algorithms}
\label{subsec:consensus}

One of the advantages of blockchain technologies is to validate the block trustfulness in a decentralized trustless environment without the necessity of the trusted third-party authority. In distributed environment, it is challenging to reach a \emph{consensus} on a newly-generated block as the consensus may be biased in favor of malicious nodes. This trustfulness validation in a decentralized environment can be achieved by \emph{consensus} algorithms. Typical consensus algorithms include proof of work (PoW), proof of stake (PoS) and practical byzantine fault tolerance (PBFT) \cite{castro1999practical}. 

Take PoW as an example. The creation of a newly-generated block is equivalent to the solution of a computationally-difficult problem. This computationally-difficult problem (\textit{aka} a puzzle) can nevertheless be verifiable without difficulty \cite{LI:FGCS2017}. Each node in the distributed peer-to-peer (P2P) network can participate in the validation procedure. The first node who solves the puzzle can append the validated block to the blockchain; this node is also called a \emph{miner}. It then broadcasts the validation results in the whole blockchain system, consequently other nodes validating and updating the new results in the blockchain. A small portion of bonus will then be given to this node as a compensation for solving the puzzle. 

\emph{Discrepancy solution.}
In a distributed system, multiple nodes may validate blocks nearly at the same time. Meanwhile, the network latency can somehow result in bifurcated (or forked) chains at the same time. To solve the discrepancy, most of existing blockchain systems typically maintain the longest chain as the valid chain because the longest chain implies the most tolerant of being compromised by adversaries. If so, a shorter chain is automatically deserted (\ie, the blue dash-line box as shown in Fig. \ref{fig:blockchain}) and the future validation work will continue on the longest chain.

\emph{Trustfulness of PoW.}
The trustfulness of PoW is based on the assumption that a majority of blockchain nodes is trustful. Generally, 51\% of computational capability is regarded as the threshold of PoW being tolerant of malicious attacks. The incentive mechanisms can encourage miners to be honest against compromising. Meanwhile, solving the puzzle typically requires extensive computing power. The probability of solving the puzzle at a miner is often proportional to the computational capability and resource of a miner \cite{MConti:CST2018}.

PoW schemes require extensive computation to solve the puzzle, thereby resulting in the extensive energy consumption. Unlike PoW, PoS requires the proof of ownership to validate the trustfulness of a block since the users with more cryptocurrencies (\ie, more stakes) are more trustful than those with fewer cryptocurrencies. In PBFT, each node who has the equal right to vote for the consensus will send its voting state to other nodes. After multiple rounds of voting procedure, the consensus reaches.

We roughly categorize typical consensus algorithms into two types: 1) Probabilistic consensus algorithms and 2) Deterministic consensus algorithms. Table \ref{tab:consensus} gives the taxonomy. Probabilistic consensus algorithms including PoW, PoS and Delegated proof of stake (DPOS) typically first save the validated block to the chain and then seek the consensus of all the nodes while deterministic consensus algorithms first consent to the block and then saved the validated block to the chain. Moreover, probabilistic consensus algorithms often result in multiple bifurcate chains and the discrepancy is solved by choosing the longest chain. In contrast, deterministic consensus algorithms solve the discrepancy through multiple rounds of communications in the overlay network. 

\begin{table}[t]
\caption{Taxonomy of typical consensus algorithms}
\centering
\renewcommand{\arraystretch}{1.5}
\begin{tabular}{m{1.2cm}|m{3.2cm}|m{3.2cm}}
\hline
& \textbf{Probabilistic Consensus} & \textbf{Deterministic Consensus} \\
\hline
\hline
Consensus procedure & Saving first and then consenting & Consenting first and then saving\\
\hline
Bifurcation (fork) & Yes & No \\
\hline
Arbitration mechanism & Choosing the longest chain when there are multiple forked chains & Voting to solve discrepancy through multiple communication-rounds \\
\hline
Adversary tolerance & $<$ 50\% computing or stakes & $<$ 1/3 voting nodes \\
\hline 
Complexity & High computational-complexity & High network-complexity \\
\hline
Examples & PoW, PoS, DPOS & PBFT and PBFT variants, Tendermint\\
\hline
\end{tabular}
\label{tab:consensus}
\end{table}

There are many attempts to improve incumbent consensus algorithms, such as Ripple \cite{chase2018analysis}, Algorand \cite{Gilad:SOSP2017}, Tendermint, proof of authority (PoA) \cite{FRYu:Access18}, proof of elapsed time (PoET) \cite{Dinh:SIGMOD2017}. Instead of choosing single consensus algorithm, there is a trend of integrating multiple consensus algorithms to fulfill the requirements from different applications. 

\subsubsection{Working flow of blockchains}
We next show how a blockchain works in an example. Take a money transfer as an example as shown in Fig. \ref{fig:blockchain-working}. Alice wants to transfer an amount of money to Bob. She first initiates the transaction at a computer through her Bitcoin wallet (\ie, Step \circled{\small 1}). The transaction includes the information such as the sender's wallet, the receiver's address and the amount of money. The transaction is essentially signed by Alice's private key and can be accessible and verifiable by other users via Alice's public key thereafter. Then the computer broadcasts the initiated transaction to other computers (or nodes) in the P2P network  (\ie, Step \circled{\small 2}). Next, a validated transaction is then appended to the end of the chain of transactions consequently forming a new block in the blockchain once a miner successfully solves the puzzle (\ie, Step \circled{\small 3}). Finally, every node saves a replica of the updated blockchain when the validated transaction is appended to the blockchain (\ie, Step \circled{\small 4}).

\subsection{Key Characteristics of Blockchain}
\label{subsec:keychar}

In summary, blockchain technologies have the following key characteristics.

\begin{itemize}

\item \emph{Decentralization.} In traditional transaction management systems, the transaction validation has been conducted through a trusted agency (\eg, a bank or government). This centralization manner inevitably results in the extra cost, the performance bottleneck and the single-point failure (SPF) at centralized service providers. In contrast, blockchain allows the transaction being validated between two peers without the authentication, jurisdiction or intervention done by the central agency, thereby reducing the service cost, mitigating the performance bottleneck, lowering the SPF risk.

\item \emph{Immutability.} A blockchain consists of a consecutively-linked chain of blocks, in which each link is essentially an inverse hash point of previous block. Any modification on the previous block invalidates all the consequently-generated blocks. Meanwhile, the root hash of the Merkle tree saves the hash of all the committed transactions. Any (even tiny) changes on any transactions generates a new Merkle root. Therefore, any falsification can be easily detected. The integration of the inverse hash point and the Merkle tree can guarantee the data integrity.

\begin{figure}[t]
\centering
\includegraphics[width=8.8cm]{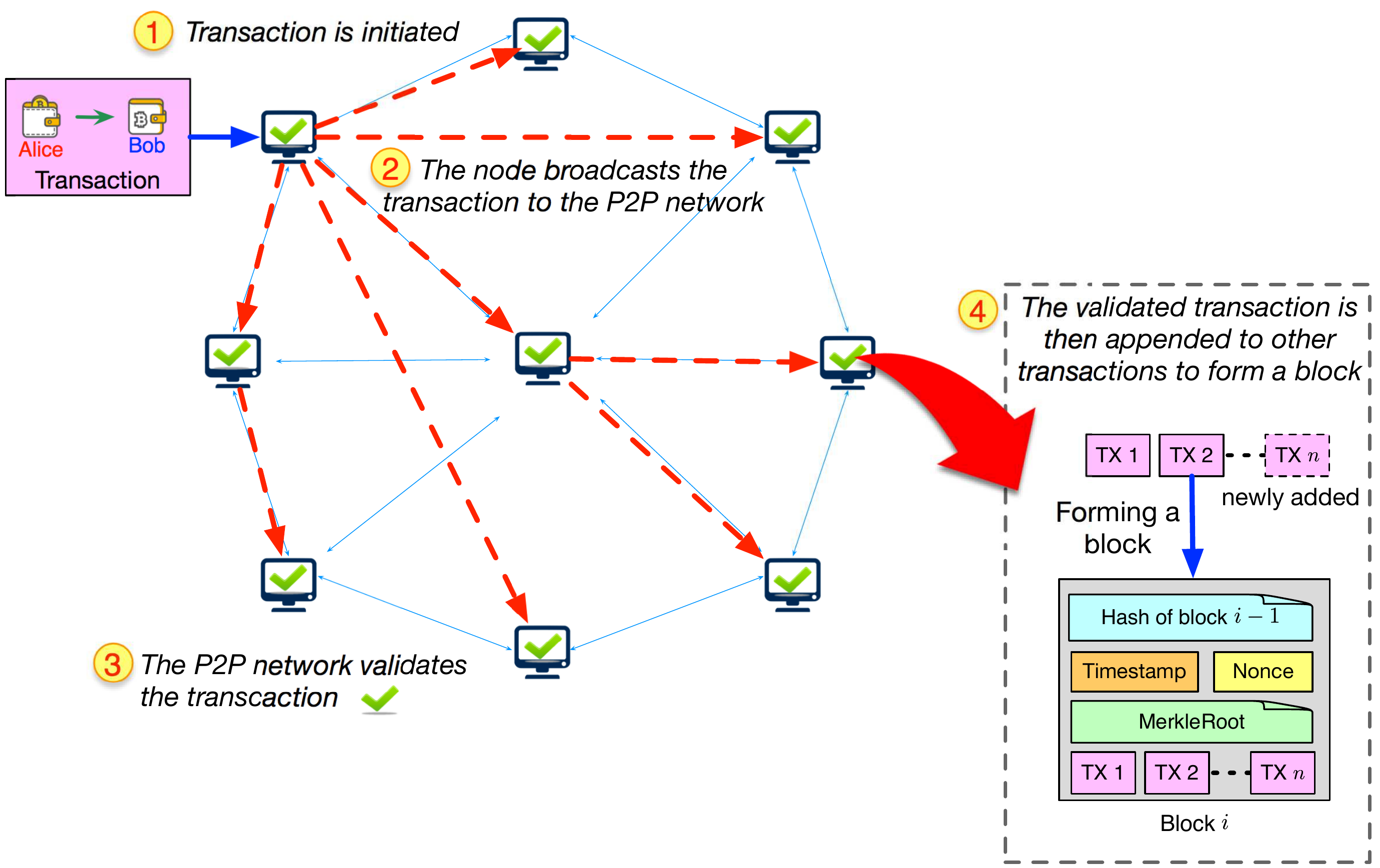}
\caption{Working flow of blockchains}
\label{fig:blockchain-working}
\end{figure}

\item \emph{Non-repudiation.} Recall the fact that the private key is used to put the signature to the transaction, which can then be accessible and verified by others via the corresponding public key. Therefore, the crytographically-signed transaction cannot be denied by the transaction initiator.

\item \emph{Transparency.} For most of public blockchain systems (like Bitcoin and Ethereum), every user can access and interact with the blockchain network with an equal right. Moreover, every new transaction is validated and saved in the blockchain, consequently being available for every user. Therefore, the blockchain data is essentially transparent to every user who can access and verify the committed transactions in the blockchain. 

\item \emph{Pseudonymity.} Despite the transparency of blockchain data, blockchain systems can preserve a certain level of the privacy via making blockchain addresses anonymous. For example, the work of \cite{Zyskind:IEEESPW15} presents an application of blockchain to preserve the privacy of personal data. However, blockchain can only preserve the privacy at a certain level since blockchain addresses are essentially traceable by inference \cite{MSAli:CST2018}. For example, it is shown in \cite{Chawathe2019} that the analysis of blockchain data can help to detect fraud and illegal transactions. Therefore, blockchain can only preserve the pseudonymity instead of full privacy. 

\item \emph{Traceability.} Each transaction saved in the blockchain is attached with a timestamp (recorded when the transaction occurs). Therefore, users can easily verify and trace the origins of historical data items after analyzing the blockchain data with corresponding timestamps. 

\end{itemize}

\subsection{Smart Contract}
\label{subsec:smart-contract}

Smart contracts are a great advance for blockchain technology \cite{ream2016upgrading}. In 1990s, smart contracts were proposed as a computerized transaction protocol that executes the contractual terms of an agreement \cite{szabo1997idea}. Contractual clauses that are embedded in smart contracts will be enforced automatically when a certain condition is satisfied (\eg, one party who breaches the contract will be punished automatically).

\begin{figure}[t]
 \centering
 \includegraphics[width=8.8cm]{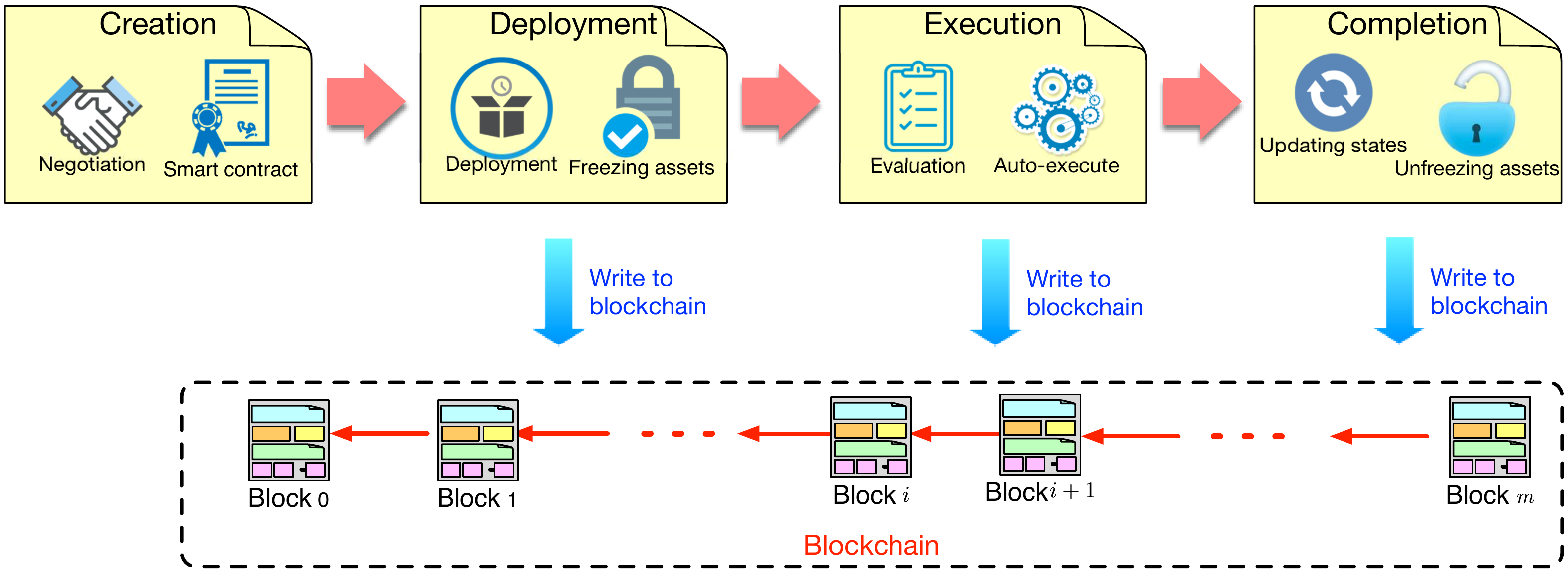}
 \caption{Life cycle of smart contracts consisting of four consecutive phases: Creation, Deployment, Execution and Completion}
 \label{fig:contract}
\end{figure}

Blockchains are enabling smart contracts. Essentially, smart contracts are implemented on top of blockchains. The approved contractual clauses are converted into executable computer programs. The logical connections between  contractual clauses have also been preserved in the form of logical flows in programs (\eg, \texttt{if-else-if} statement). The execution of each contract statement is recorded as an immutable transaction stored in the blockchain. Smart contracts guarantee appropriate access control and contract enforcement. In particular, developers can assign access permission for each function in the contract. Contract enforcement ensures that the contract execution is deterministic. Once any conditions in a smart contract are satisfied, the triggered statement will automatically execute the corresponding function in a predictable manner. For example, Alice and Bob agree on the penalty of violating the contract. If Bob breaches the contract, the corresponding penalty (as specified in the contract) will be automatically paid from Bob's deposit.

The whole life cycle of smart contracts consists of four consecutive phases~\cite{ZZheng:FGCS20} as illustrated in Fig. \ref{fig:contract}:
\begin{enumerate}
\item  \textit{Creation} of smart contracts. Several involved parties first negotiate on the obligations, rights and prohibitions on contracts. After multiple rounds of discussions and negotiations, an agreement can reach. Lawyers or counselors will help parties to draft an initial contractual agreement. Software engineers then convert this agreement written in natural languages into a smart contract written in computer languages including declarative language and logic-based rule language \cite{Idelberger:2016}. Similar to the development of computer software, the procedure of the smart contract conversion is composed of design, implementation and validation (\ie, testing). It is worth mentioning that the creation of smart contracts is an iterative process involving with multiple rounds of negotiations and iterations. Meanwhile, it is also involved with multiple parties, such as stakeholders, lawyers and software engineers.

\item  \textit{Deployment} of smart contracts. The validated smart contracts can then be deployed to platforms on top of blockchains. Contracts stored on the blockchains cannot be modified due to the immutability of blockchains. Any emendation requires the creation of a new contract. Once the smart contracts are deployed on blockchains, all the parties can access the contracts through the blockchains. Moreover, digital assets of both involved parties in the smart contract are locked via freezing the corresponding digital wallets \cite{Sillaber2017}. For example, the coin transfers (either incoming or outgoing) on the wallets relevant to the contract are blocked. Meanwhile, the parties can be identified by their digital wallets. 

\item  \textit{Execution} of smart contracts. After the deployment of smart contracts, the contractual clauses have been monitored and evaluated. Once the contractual conditions reach (\eg, product reception), the contractual procedures (or functions) will be automatically executed. It is worth noting that a smart contract consisting of a number of declarative statements with logical connections. When a condition is triggered, the corresponding statement will be automatically executed, consequently a transaction being executed and validated by miners in the blockchains \cite{koulu2016blockchains}. The committed transactions and the updated states have been stored on the blockchains thereafter. 

\item  \textit{Completion} of smart contracts. After a smart contract has been executed, new states of all involved parties are updated. Accordingly, the transactions during the execution of the smart contracts as well as the updated states are stored in blockchains. Meanwhile, the digital assets have been transferred from one party to another party (\eg, money transfer from the buyer to the supplier). Consequently, digital assets of involved parties have been unlocked. The smart contract has completed the whole life cycle.
\end{enumerate}

It is worth mentioning that during deployment, execution and completion of a smart contract, a sequence of transactions has been executed (each corresponding to a statement in the smart contract) and stored in the blockchain. Therefore, all the three phases need to write data to the blockchain as shown in Fig. \ref{fig:contract}.

\subsection{Taxonomy of Blockchain Systems}
\label{subsec:taxonomy}

\begin{table}[t]
\caption{Comparisons of Blockchain systems}
\centering
\renewcommand{\arraystretch}{1.5}
\begin{tabular}{m{1.7cm} |m{1.5cm} | m{1.5cm} | m{2.2cm} }
\hline
 &\bf{Public} & \bf{Private} & \bf{Consortium} \\ \hline\hline

Decentralization & Decentralized & Centralized & Partially Decentralized\\ 

Immutability & Immutable & Alterable & Partially Immutable\\ 

Non-repudiation & Non-refusable & Refusable & Partially Refusable\\

Transparency & Transparent & Opaque & Partially Transparent \\ 

Traceability & Traceable & Traceable   & Partially Traceable   \\ 
\hline

Scalability & Poor & Superior & Good \\ 
\hline

Flexibility & Poor & Superior & Good \\ 
\hline

Permission & Permissionless & Permissioned & Permissioned \\ 
\hline
Consensus & PoW, PoS & Ripple & PBFT, PoA, PoET \\
\hline 
Examples & Bitcoin \cite{nakamoto2008bitcoin}, Ethereum\cite{Ethereum} & GemOS \cite{GemOS}, Multichain \cite{MultiChain}  & Hyperledger \cite{hybperledge:2015} Ethereum\cite{Ethereum} \\
\hline

\end{tabular}
\label{tab:comp-blockchains}
\end{table}

We classify blockchain systems into three types: 1) public blockchains, 2) private blockchains and 3) consortium (or community) blockchains \cite{Xu:ICSA2017}. Most digital currencies such as BTC (\ie, the ticker symbol of Bitcoin cryptocurrency) and ETH (\ie, the ticker symbol of Ethereum cryptocurrency) are implemented on public blockchains, thereby being accessible by anyone in the P2P network. Differently, private blockchains can be managed or controlled by a single organization while consortium blockchains sit in limbo between public and private blockchains. Table \ref{tab:comp-blockchains} presents a comparison of three types of blockchains.

In particular, we summary the comparison among public, private and consortium blockchains in the following aspects.
\begin{itemize}

\item \emph{Key characteristics.} Public blockchains are fully-decentralized while private and consortium blockchains are
partially decentralized or fully controlled by a single group or multiple groups. Moreover, it is nearly impossible to tamper transactions in public blockchains as every node keeps a replica of the blockchain (containing all the transactions) while the dominant organization or multiple parties of consortium and private blockchains can modify the blockchain. Similarly, public blockchains can fully ensure the non-repudiation, transparency and traceability of transactions while private and consortium blockchains cannot or can only partially ensure these properties.

\item \emph{Scalability.} Although public blockchains can guarantee the decentralization, immutability, transparency, non-repudiation and traceability, the merits are obtained in the cost of low transaction-validation rate, high latency and extra storage space consumption, consequently limiting the scalability of public blockchains. Compared with public blockchains, private and consortium blockchains have a better scalability since blockchains are fully controlled by a single group or multiple organizations and the consensus can be easily reached. 

\item \emph{Flexibility.} Similarly, public blockchains have the less flexibility than private and consortium blockchains since configurations of private and consortium blockchains are more adjustable.

\item \emph{Permission.} Permission refers to consent or authorization to access the blockchains. In public blockchains, public participation is allowed, thereby being permissionless. However, private and consortium blockchains can allow one or more users to access and interact with blockchains with different permission levels. For example, some users can only read the blockchain data while others can either read or initiate transactions. 

\item \emph{Consensus.} Public blockchains usually use PoW and PoS as the consensus algorithms, which are Byzantine-failure tolerant while resulting in extensive resource consumption. Private blockchains can easily achieve the consensus among the authenticated users. Typical consensus algorithms used for private blockchains include PBFT, PoA and PoET. Moreover, consortium blockchains are a hybrid type of public blockchains and private blockchains. In particular, Ripple \cite{chase2018analysis} is a variant of PBFT typically used for consortium blockchains.

\begin{figure*}[t]
\centering
\subfigure[Blockchain-composite layer]{
\includegraphics[width=8.7cm]{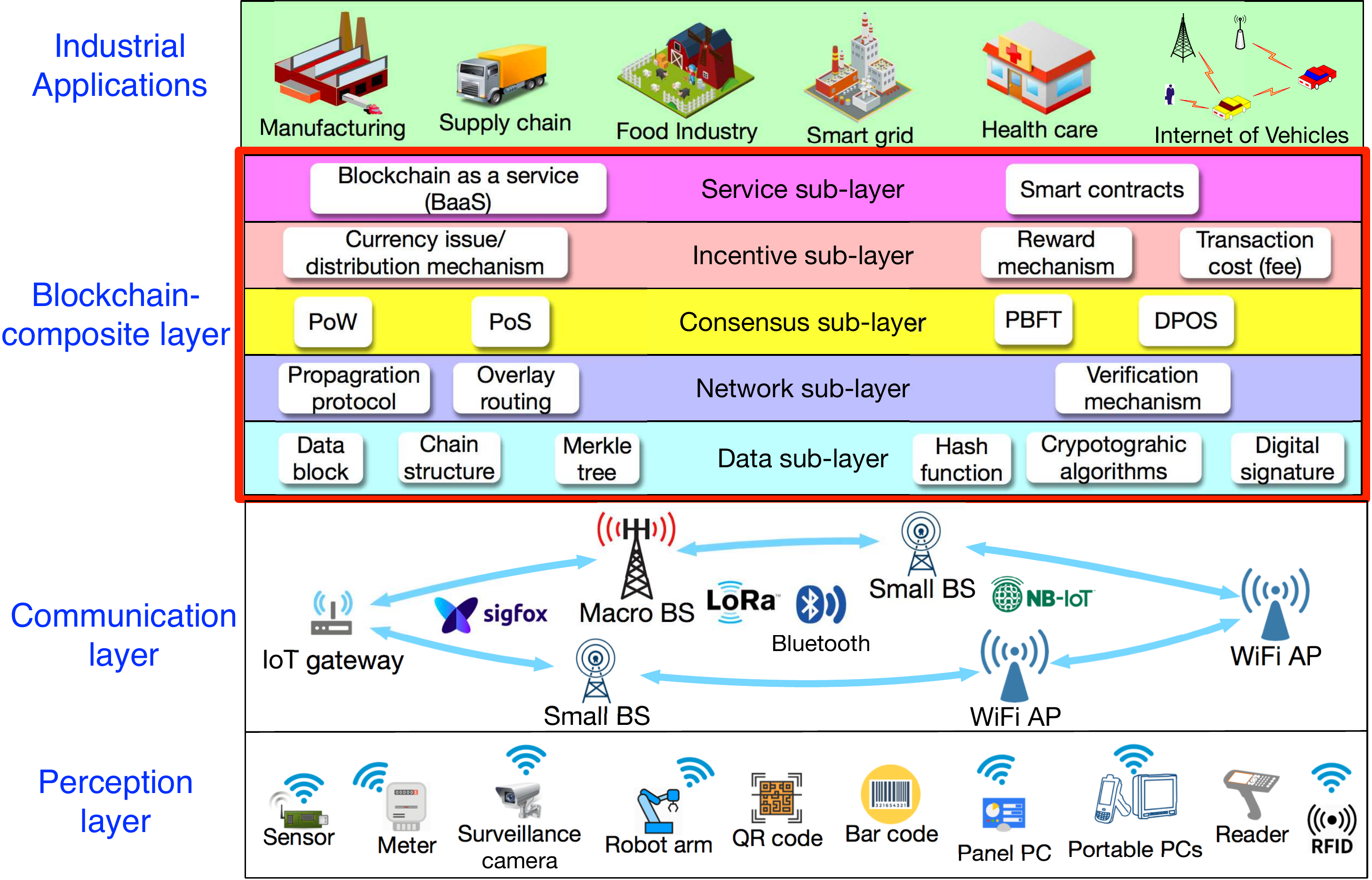}
\label{fig:ibot-layer}}
\subfigure[P2P overlay network and blockchain node architecture]{
\includegraphics[width=8.7cm]{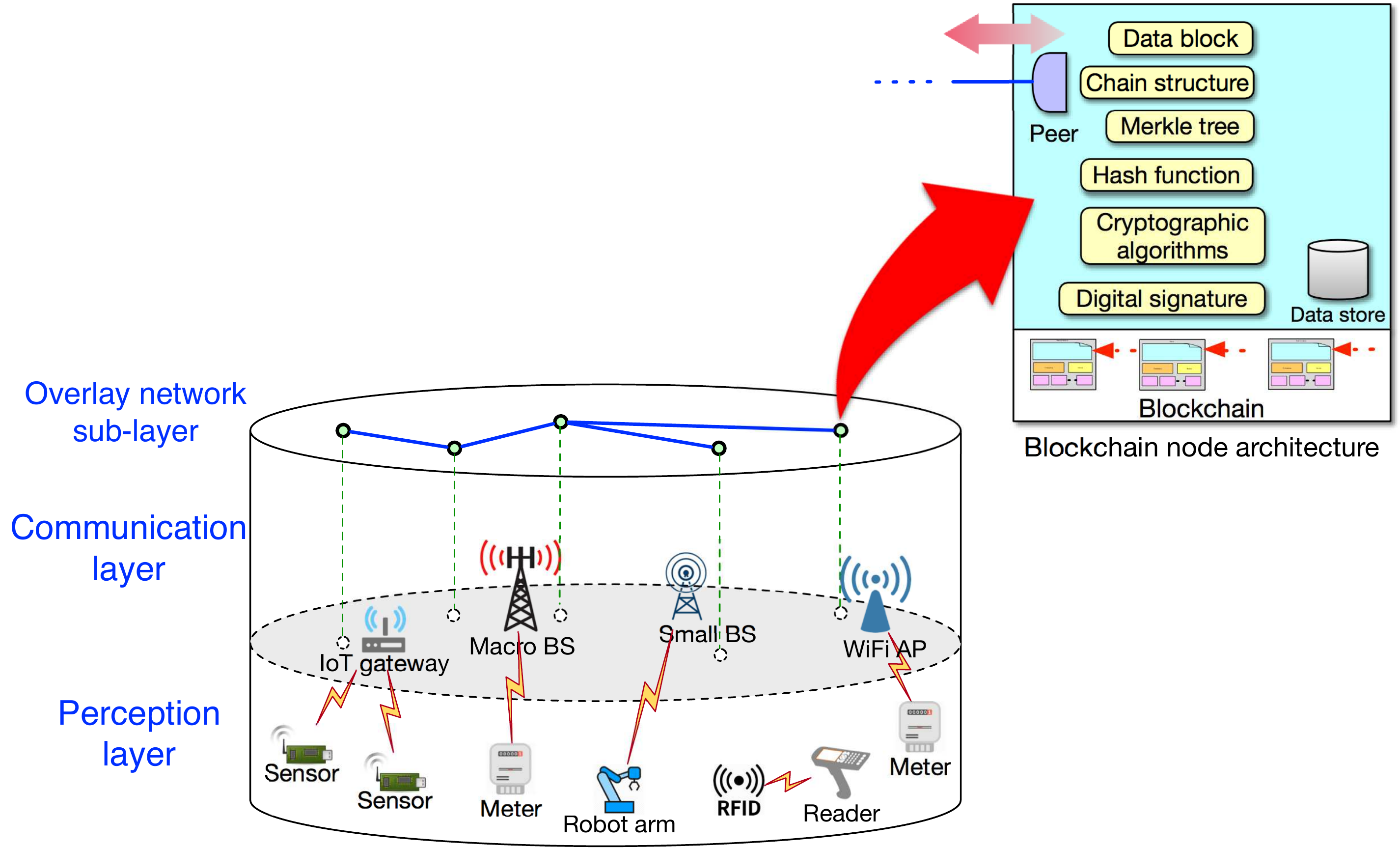}
\label{fig:ibot-deployment}}
\caption{Overview of BCoT architecture}
\label{fig:ibot}
\end{figure*}

\item \emph{Exemplary platforms.} Bitcoin \cite{nakamoto2008bitcoin} and Ethereum\cite{Ethereum} are two typical public blockchain platforms, which are mainly used for digital currency. With regard to private blockchains, GemOS \cite{GemOS} is a private blockchain platform for healthcare and supply chain. In addition, MultiChain \cite{MultiChain} is an open source platform granting the implementation of private blockchains. As for consortium blockchains, Hyperledger \cite{hybperledge:2015} is developing business consortium blockchain frameworks. Moreover, Ethereum also provides tools for building consortium blockchains \cite{consortium}.
\end{itemize}

%
%
%
%
%
%
%


\section{Convergence of Blockchain and IoT} 
\label{sec:IBoT}

In this section, we first discuss the opportunities of integrating blockchain with IoT in Section \ref{subsec:opp}. We then present the architecture of the integration of blockchain and IoT (namely BCoT) in Section \ref{subsec:architecture}. We next discuss the deployment issues on BCoT in Section \ref{subsec:deployment}.

\subsection{Opportunities of integrating blockchain with IoT}
\label{subsec:opp}

As summarized in Section \ref{subsec:challenges-IIoT}, IoT systems are facing many challenges such as heterogeneity of IoT systems, poor interoperability, resource constraints of IoT devices, privacy and security vulnerabilities. Blockchain technologies can complement IoT systems with the enhanced interoperability and the improved privacy and security. Moreover, blockchain can also enhance the reliability and scalability of IoT systems \cite{REYNA:2018}. In short, we name such integration of blockchain with IoT as BCoT. BCoT has the following potential benefits in contrast to incumbent IoT systems.
\begin{itemize}
\item \emph{Enhanced interoperability} of IoT systems. Blockchain can essentially improve the interoperability of IoT systems via transforming and storing IoT data into blockchains. During this procedure, heterogeneous types of IoT data are converted, processed, extracted, compressed and finally stored in blockchains. Moreover, the interoperability also exhibits in easily passing through different types of fragmented networks since blockchains are established on top of the P2P overlay network that supports universal internet access. 
 
\item \emph{Improved security} of IoT systems. On one hand, IoT data can be secured by blockchains since they are stored as blockchain transactions which are encrypted and digitally-signed by cryptographic keys (\eg, elliptic curve digital signature algorithm \cite{Johnson:2001}). Moreover, the integration of IoT systems with blockchain technologies (like smart contracts) can help to improve the security of IoT systems by automatically-updating IoT device firmwares to remedy vulnerable breaches thereby improving the system security \cite{christidis2016blockchains}.

\item \emph{Traceability} and \emph{Reliability} of IoT data. Blockchain data can be identified and verified anywhere and anytime. Meanwhile, all the historical transactions stored in the blockchains are \emph{traceable}. For example, the work of \cite{QLu:IEEESoftware2017} has developed a blockchain-based product traceability system, which provide suppliers and retailers with traceable services. In this manner, the quality and originality of the products can be inspected and verified. Moreover, the immutability of blockchains also assures the reliability of IoT data since it is nearly impossible to alter or falsify any transactions stored in blockchains. 

\item \emph{Autonomic} interactions of IoT systems. Blockchain technologies can grant IoT devices or subsystems to interact with each other automatically. For example, the work of \cite{zhang2015iot} proposes Distributed autonomous Corporations (DACs) to automate transactions, in which there are no traditional roles like governments or companies involved with the payment. Being implemented by smart contracts, DACs can work automatically without human intervention consequently saving the cost.
		
\end{itemize}

\subsection{Architecture of Blockchain of Things}
\label{subsec:architecture}

We propose the architecture of BCoT as shown in Fig. \ref{fig:ibot}. In this architecture, the blockchain-composite layer plays as a middleware between IoT and industrial applications. This design has two merits: 1) offering an abstraction from the lower layers in IoT and 2) providing users with blockchain-based services. In particular, the blockchain-composite layer hides the heterogeneity of lower layers (like perception layer and communication layer in IoT). On the other hand, the blockchain-composite layer offers a number of blockchain-based services, which are essentially application programming interfaces (APIs) to support various industrial applications. As a result, the difficulty of developing industrial applications can also be lowered down due to the abstraction achieved by the blockchain-composite layer. 

\begin{figure*}[t]
 \centering
 \includegraphics[width=14.8cm]{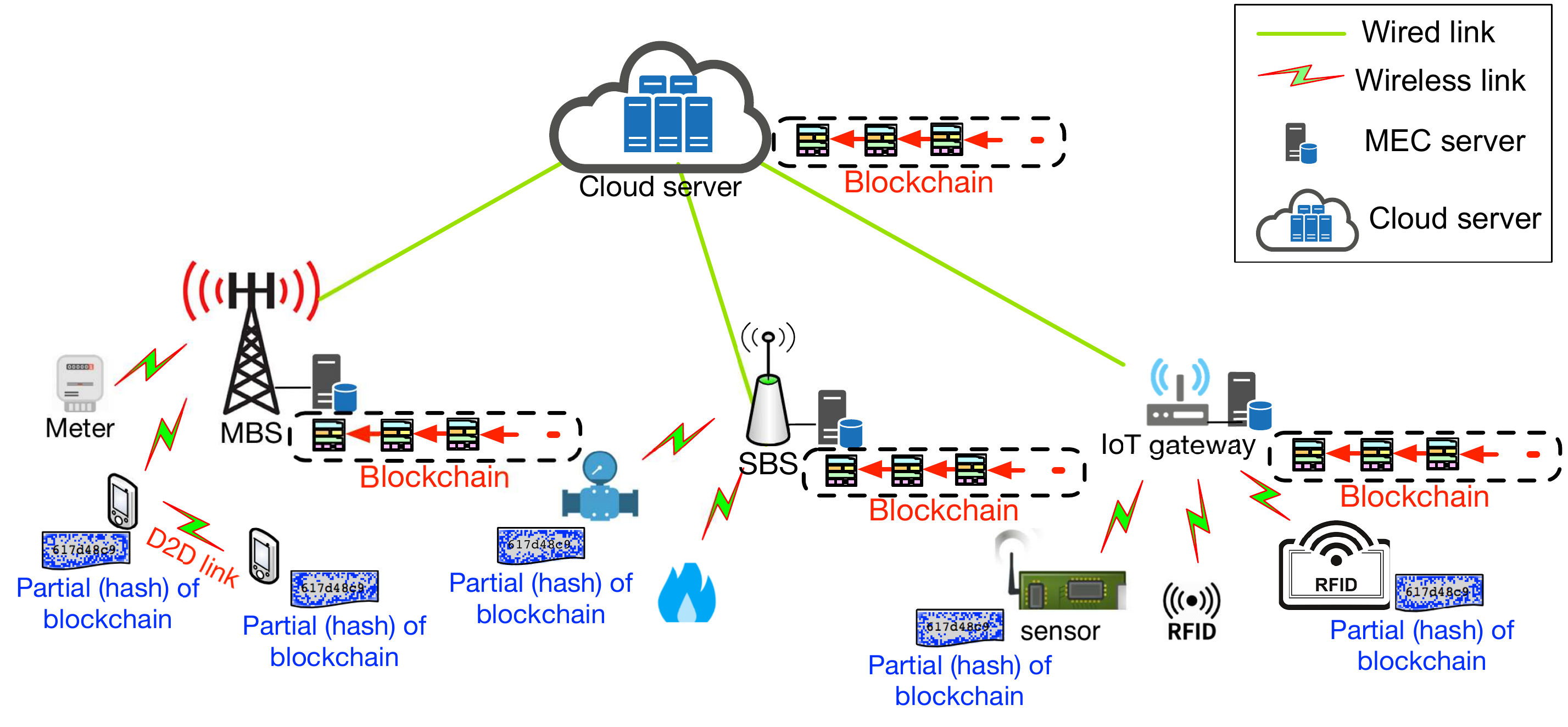}
 \caption{Deployment scenario of BCoT}
 \label{fig:IoTEdgeCloud}
\end{figure*}

In particular, the blockchain-composite layer consists of 5 sub-layers as shown in Fig. \ref{fig:ibot-layer} (from bottom to up): 
\begin{enumerate}
\item \emph{Data sub-layer} collects the IoT data from the lower layers (\eg, perception layer) and wraps up the encrypted data with digital signature via asymmetric cryptographic algorithms and hash functions. These consecutively-connected data blocks then form the blockchain after the distributed validation. Different blockchain platforms may choose different cryptographic algorithms and hash functions. For example, Bitcoin blockchain chooses SHA-256 as the hash function and elliptic curve digital signature algorithm (ECDSA) as the signature algorithm.

\item \emph{Network sub-layer} is essentially an overlay P2P network running on top of the communication layer. The overlay network consists of either virtual or physical links connecting nodes in the underlying communication networks (\ie, wired/wireless communication networks). One node only simply broadcasts the block of transactions to its connected peers. Once receiving the block of transactions, other peers will verify it locally. If it is valid, the block will be further propagated to other nodes through the overlay network.

\item \emph{Consensus sub-layer} is mainly involved with the distributed consensus for the trustfulness of a block. The consensus can be achieved by various consensus algorithms like PoW, PoS, PBFT and DPOS (as explained in Section \ref{subsec:consensus}). It is worth mentioning that block propagation mechanisms (such as relay network propagration and advertisement-based propagation \cite{LI:FGCS2017}) are the prerequisite for the distributed consensus protocols. 

\item \emph{Incentive sub-layer} is responsible for the following tasks: 1) digital currency issuing, 2) digital currency distribution, 3) designing reward mechanism (especially for miners), 4) handling transaction cost, etc. In particular, it is important to design appropriate monetary policy of digital currency (\ie, money creation and distribution), distribute rewards to participants who contribute to distributed consensus (\ie, mining).

\item \emph{Service sub-layer} provides users with blockchain-based services for various industrial sectors include manufacturing, logistics, supply chains, food industry and utilities. The blockchain as a service (BaaS) can be achieved by smart contracts, which can be automatically triggered when a special event occurs. For example, a payment contract is automatically executed when a product is well received by a consumer.

\end{enumerate}

It is worth mentioning that the network sub-layer that is established on top of the communication layer is the abstraction of underneath communication networks, consequently offering a universal network access across different networks as shown in Fig. \ref{fig:ibot-deployment}. Fig. \ref{fig:ibot-deployment} also shows the architecture of a blockchain node, which essentially includes blockchain data and other elements in the data sub-layer. 

\subsection{Deployment of BCoT}
\label{subsec:deployment}

The realistic deployment of BCoT is of great importance. However, due to the constraints of IoT devices, it is challenging to store the whole blockchain at IoT devices. In particular, there are two modes to store the blockchain data \cite{REYNA:2018}: i) \emph{full storage}, in which the entire blockchain is stored, ii) \emph{partial storage}, in which only a subset of data blocks are stored locally. Accordingly, we name the nodes with full storage of blockchain data as \emph{full nodes} and the nodes with partial storage of blockchain data as \emph{lightweight nodes}. In practice, a full node can be a cloud server or an edge server  with adequate computing resources since it requires a large storage space to save the entire blockchain (\eg, the whole Bitcoin blockchain occupies nearly 185 GB at the end of September 2018 according to the statistic report\footnote{https://www.statista.com/statistics/647523/worldwide-bitcoin-blockchain-size/}) and strong computing capability of solving consensus puzzles (\ie, mining). On the other hand, resource-constrained IoT devices (\eg, sensors, IoT objects) can be lightweight nodes that can validate the trustfulness of a transaction without downloading or saving the whole blockchain (\ie, only saving partial blockchain data such as hash values). It is worth mentioning that the lightweight nodes highly rely on the full nodes. 

Fig. \ref{fig:IoTEdgeCloud} presents a possible deployment scenario of BCoT, in which cloud servers and edge servers may store the whole blockchain (or partial blockchain) data while IoT devices may only save the particial blockchain data. In addition to the deployment of BCoT, there are also several possible interaction manners between IoT and blockchain \cite{MSAli:CST2018}: (i) direct interaction between IoT and blockchain, in which IoT devices can directly access blockchain data saved at edge servers co-located with IoT gateways, Macro Base Stations (MBS) or Small BS; (ii) direct interaction between IoT nodes, in which IoT nodes can directly exchange/access partial blockchain data via D2D links; (iii) hybrid interaction of cloud and edge servers with IoT devices, in which IoT devices can interact with blockchain data through edge/cloud servers. 

There are several initiatives addressing the configuration and initialization of blockchain at edge servers or at IoT devices. For example, Raspnode\footnote{http://raspnode.com/} is a project mainly for installing Bitcoin and other blockchains at Raspberry Pi micro computers. EthArmbian\footnote{http://raspnode.com/} offers the customized Ubuntu Linux image for ARM devices, each of which can serve as an Ethereum node. Despite these initiatives, most of IoT devices are still lightweight nodes due to the limited storage. 

\section{Blockchain for 5G Beyond in IoT} 
\label{sec:5G}
\begin{figure*}[t]
 \centering
 \includegraphics[width=14.8cm]{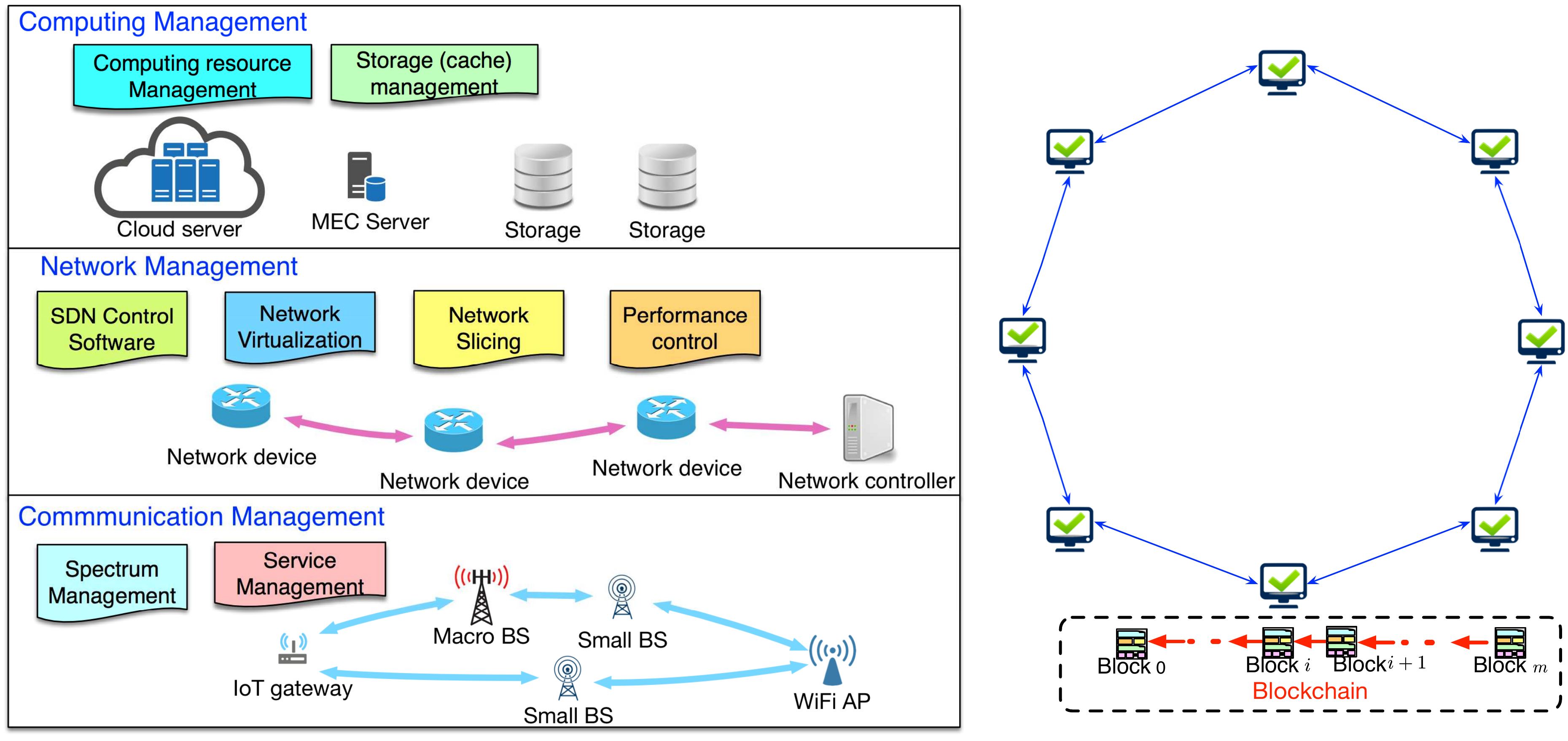}
 \caption{Blockchain for 5G Beyond Networks in IoT}
 \label{fig:5Gbeyond}
\end{figure*}

Although blockchain technology is promising to IoT, there are still many research issues to be addressed before the integration of blockchain with IoT, especially for the next-generation networks (\ie, 5G-beyond or 6G networks), which play a critical role in constructing the infrastructure for blockchains. Fig. \ref{fig:5Gbeyond} illustrates the potentials brought by blockchain to 5G-beyond networks in the perspectives from communications, network management and computing management. We explain them in details as follows.

\subsection{Blockchain for communications}

The growing demands of mobile data traffic are driving the more efficient resource management in the fifth generation (5G) communication systems. For example, radio spectrum is one of the most important resources \cite{MASSARO:2017}.  Radio spectrum management typically includes spectrum auction and spectrum sharing. It is shown in the latest speech \cite{FCC:whitepaper18} given by Federal Communications Commission (FCC) commissioner J. Rosenworcel that blockchain technology could be used to achieve the dynamic and secure spectrum management in 5G and 5G beyond (\emph{aka} 6G) communication systems \cite{Saracco:6G18,Gatherer:6G18}. The benefits of using blockchains for 5G-beyond networks lie in the secure and traceable transaction-management without the necessity of a central intermediary, consequently saving the management cost. Ref. \cite{Seppo:2018} gives several use cases to illustrate that using blockchain technology can benefit radio spectrum sharing in terms of trustfulness, consensus and cost reduction. Moreover, Kotobi and Bilen \cite{Kotobi:VTM2108} put forth a blockchain-based protocol to secure spectrum sharing between primary users and cognitive users in wireless communication systems. In addition, blockchain may potentially help to share link conditions to multiple IoT nodes with privacy preservation consequently improving spectral efficiency via traffic optimization \cite{Kure:IoTJ19}.

In addition to the radio spectrum management, blockchains also have the potentials to provide users with the improved mobile services. For example, 5G networks typically consist of a number of fragmented heterogeneous networks. Blockchains that are built on top of the network layer can help to integrate different networks with the provision of seamless access between different networks. Moreover, smart contracts can automate the procedure of provisions and agreements between network operators and subscribers while operational cost can be greatly saved \cite{Huawei:whitepaper17}. The work of \cite{Sheng:ICBC2018} also shows that a blockchain-based system can help operating nodes to improve their operational and service capabilities. In the future, the synthesis of blockchains and big data analytics can help service providers to extract valuable insights from transactions of subscribers and offer the better services for users \cite{hndai:BDAWireless2019}.

\subsection{Blockchain for network management}

Recently, software defined networking (SDN) technology can bestow the flexibility and scalability for distributed IoT \cite{Bera:IoTJ2017}. However, it is shown in \cite{Kalkan:ComMag2017} that the centralization of SDN can also result in the single-point-of-failure. Moreover, incumbent SDN devices (such as gateways) are also incapable of conducting computational-intensive analysis on data traffic. The integration of blockchain technology with SDN can overcome the disadvantages of SDN. For example, the work of \cite{Sharma:ComMag2017} proposes a secure blockchain-based SDN framework for IoT. In particular, a blockchain-based scheme has been developed to update the flow rule table in a secure way without the necessity of the intermediary. In addition, blockchain can also help to secure the network management of network function visualization (NFV). In particular, it is shown in \cite{Alvarenga:NOMS18} that the integration of blockchain with NFV can ensure that the configuration of NFV is immutable, auditable, non-repudiable, consistent and anonymous. A prototype of the proposed architecture was also developed and implemented in this work. 

In addition to SDN and NFV, the appearance of network slicing technologies \cite{Afolabi:CST2018} brings the agility and flexibility of networks to support different functional and performance requirements. As mentioned in Section \ref{sec:IBoT}, different industrial sectors have diverse application demands on blockchains. For example, a single blockchain is typically used in digital-currency like applications while an enterprise may maintain several blockchains to serve for different purposes. In particular, four isolated blockchains are dedicate to Enterprise resource planning (ERP), Product Lifecycle Management (PLM), Manufacturing execution systems (MES) and Customer Relationship Management (CRM), respectively \cite{Esposito:IEEECloudComp2016}. Network slicing can essentially offer a solution to the diverse demands of blockchain applications in mobile edge computing. For example, each of network instances can be created for the provision of a specific blockchain service on top of network slicing and network visualization. However, it is necessary to optimize and allocate both network and computing resources to fulfill the diverse demands in the composite environment of mobile edge computing and cloud computing. Moreover, the integration of blockchain and network slicing technologies can also support the reliable content sharing in content-centric networks (CCNs) \cite{Ortega:VTM2018} and privacy preservation in data sharing in 5G networks \cite{Fan:IETCom2018}.

\subsection{Blockchain for computing management}

Due to the resource constraints of IoT devices, massive IoT data has been typically uploaded to remote cloud servers for further processing. However, the pure cloud-based computing paradigm also causes the network traffic bottlenecks, long latency, context unawareness and privacy exposure \cite{CChen:IEEENet2018}, thereby limiting the scalability of IoT. Recently, Mobile Edge Computing (MEC) \cite{Abbas:IoTJ2018} is becoming a crucial complement to cloud computing by offloading computing tasks from distant cloud servers to MEC servers typically installed at IoT gateways, WiFi APs, Macro BS and Small BS, which are close to users. In this manner, the context-aware, latency-critical and less-computing-intensive tasks can be migrated from remote cloud servers to local MEC servers, thereby improving the response, privacy-preservation and context-awareness. 

Blockchain technology has been applied in a variety of fields due to its capability of establishing trust in a decentralized fashion. There are still a number of issues needed to be solved before MEC can be used in BCoT \cite{ZXiong:ComMag2018}. In contrast to cloud servers with strong computing capability and extensive storage space, mobile edge servers usually have inferior capability. Moreover, mobile edge servers are heterogeneous in terms of computing capability, main memory, storage space and network connection. As a result, mobile edge servers cannot accommodate the computational demands alone. For example, a mobile edge server may not be able to solve the consensus puzzle in blockchains while a cloud server can serve for this goal. Therefore, it is worthwhile to investigate the orchestration of mobile edge computing and cloud computing for the provision of blockchain services \cite{MLiu:TVT18}. 

\subsection{Orchestration of cloud and edge computing with blockchain}

During the orchestration of cloud and edge computing with blockchain, there are several challenges including computational task offloading and incentivizing resource sharing. 

Offloading the computational tasks to edge servers can significantly reduce the delay. Therefore, it is crucial to conduct edge-cloud interoperation \cite{Yang:NetMag2017}. Nevertheless, it can cause a performance bottleneck and a single-point-of-failure if all the nodes offload their tasks to the same MEC server. The work of \cite{YDai:IoTJ19} presents an offloading method with consideration of load balancing among multiple MEC servers. Meanwhile, it is worthwhile to investigate how to incentivize both edge severs and cloud servers. For example, \cite{ZZhou:TVT19} presents a contract-match approach to allocate computational resource and assign tasks while incentivizing edge severs and cloud servers effectively. Moreover, it is challenging to design an optimal solution to the offloading tasks with consideration of spectrum, computation and energy consumption together. The work of \cite{YYang:IoTJ18} essentially provides a solution to optimize the offloading energy consumption with consideration of feasible modulation schemes and tasks scheduling. However, most of existing studies only consider a task is either done at an edge sever or at a cloud. In realistic application, a task can be partitioned into multiple sub-tasks with task dependency and those sub-tasks can be either executed at the edge server or at the cloud server. It is worthwhile to investigate the task partition with consideration of sub-task dependency in blockchains in the future.

\section{Applications of Blockchain of Things} 
\label{sec:IBoTapp}

\begin{figure}[t]
\centering
\includegraphics[width=9.1cm]{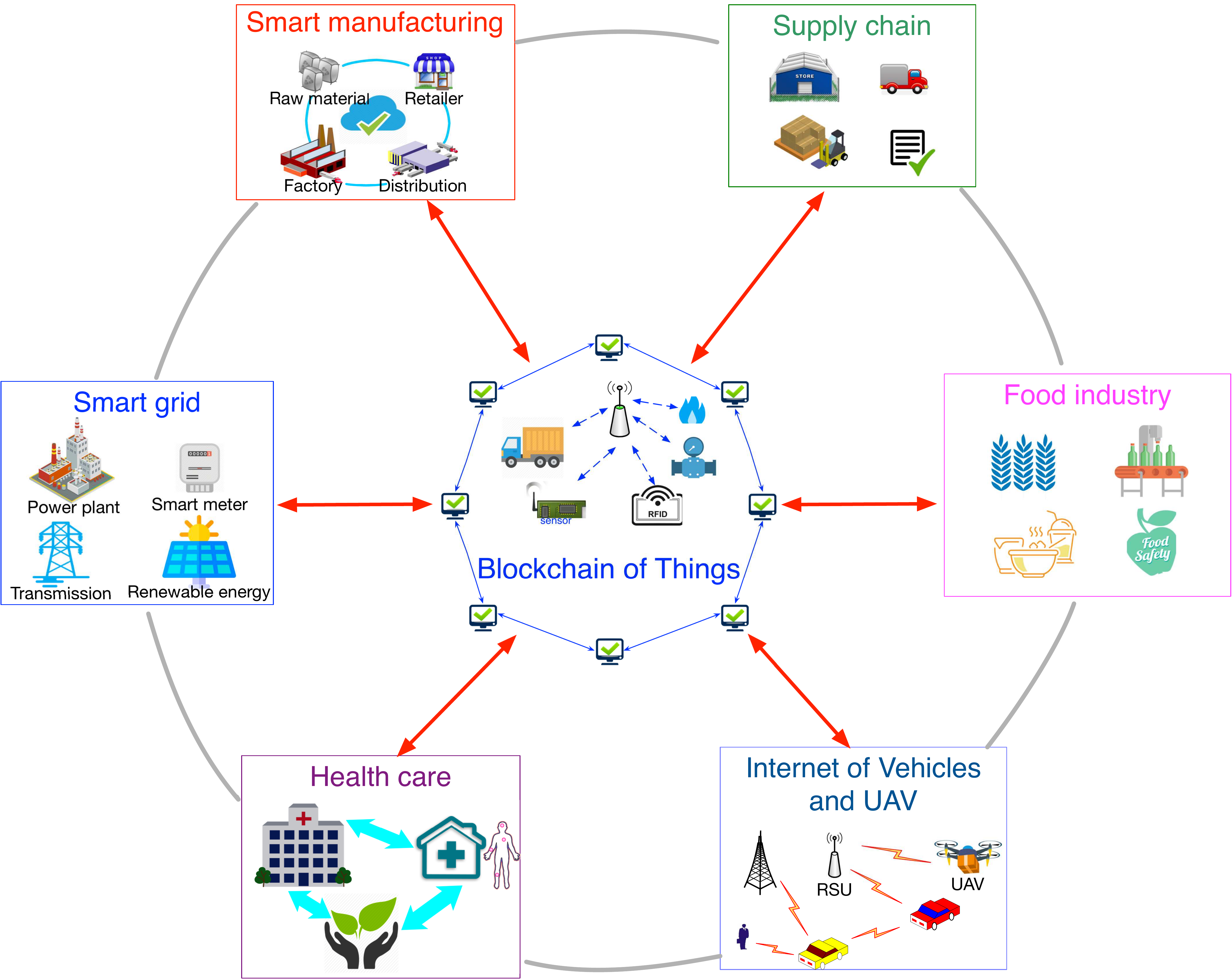}
\caption{Applications of Blockchain of Things}
\label{fig:applications}
\end{figure}

There is a growing trend in applying blockchain in IoT since blockchain technologies can help to overcome the challenges of IoT. We then provide an overview of the applications of BCoT. It is worth mentioning that there is a wide diversity of applications of blockchains (ranging from smart manufacturing to internet of vehicles and unmanned aerial vehicles). In this paper, we mainly focus on the industrial applications of BCoT. We roughly categorize the applications of BCoT into six types as shown in Fig. \ref{fig:applications}.


\subsection{Smart manufacturing}
The manufacturing industry is experiencing an upgrading from automated manufacturing to ``smart manufacturing'' \cite{Kusiak:IJPR2018}. Big data analytics on manufacturing data plays an important role during this upgrading process. Massive data is generated during every phase of the product life cycle consisting of product designing, raw material supply, manufacturing, distribution, retail and after-sales service. However, the manufacturing data is highly fragmented, consequently leading to the difficulty in data aggregation and data analytics. BCoT can address the interoperability issue by interconnecting IoT systems via P2P network and allowing data sharing across industrial sectors. For example, several distributed blockchains can be constructed to serve for different sectors and each blockchain is serving for a sector or more than one sector.

BCoT can also improve the security of smart manufacturing. One of major bottlenecks limiting the upgrading of factories is that the IoT systems have been maintained in a centralized way. For example, IoT firmware needs to be upgraded regularly to remedy security breaches. However, most of the firmware updates are downloaded from a central server and then are manually installed at IoT devices. It is expensive and in-efficient to install and upgrade the firmware updates in distributed IoT. The work of \cite{christidis2016blockchains} presents an automatic firmware upgrading solution based on smart contract and blockchains. In particular, smart contracts describing the firmware upgrading manners (\eg, when and where to upgrade firmwares) are deployed across the whole industrial network. Devices can then download and install the firmware hashes via smart contracts being  automatically executed. As a result, the security maintenance cost can be greatly saved. In addition, a decentralized blockchain-based automatic production platform was proposed in \cite{JWan:TII19} to offer a better security and privacy protection than conventional centralized architecture. In addition, a blockchain-based mobile crowdsensing system was proposed to solve the incentive issue with data quality assurance in smart manufacturing \cite{JHuang:TII20}.

\subsection{Supply chain management}

A product often consists of multiple parts provided by different manufacturers across countries. However, some forged (or low-quality) parts may seep into the supply chain. It is quite expensive to apply anti-fraud technologies in every part of a product. The integration of blockchain and IoT can solve this problem. In particular, every part will be associated with a unique ID with the creation. Meanwhile, an immutable timestamp is also attached with this ID. The identification of every part can then be saved into a blockchain, which is tamper-resistant and traceable. For example, the work of \cite{Konstantinidis:2018} shows that the part ownership of a product can be authenticated through a blockchain-based system. Moreover, the work of \cite{Kim:ISAFM2018} presents a traceability ontology with the integration of IoT and blockchain technologies based on Ethereum blockchain platform. The proposed framework has demonstrated to guarantee data provenance of supply chain.

On the other hand, BCoT can also be used to reduce the costs in after-sale services in the supply chain management. The work of \cite{tapscott2017blockchain} shows a user case of a motor insurance, in which the settlement of claims can be automated via smart contracts based on blockchains, thereby improving the efficiency and reducing the claim-processing time. Moreover, it is shown in \cite{Kshetri:2018} that integrating blockchain with IoT can help to reduce the cost, fasten the speed and reduce the risk in the supply chain management. Furthermore, a blockchain-based Machine Learning platform \cite{ZLi:TII19} was proposed to secure the data sharing among different enterprises to improve the quality of customer service.

\subsection{Food industry}

BCoT can enhance the visibility of the product life cycle especially in food industry. In particular, the traceability of food products is a necessity to ensure food safety. However, it is challenging for the incumbent IoT to guarantee the food traceability in the whole food supply chain \cite{DTse:IEEM2017}. For example, a food company may be provisioned by a number of suppliers. The traceability requires digitizing the information of raw materials from sources to every sector of food manufacturing. During this procedure, blockchain technologies can ensure the traceability and the provenance of food industry data. 

There are several proposals in this aspect. For example, the work of \cite{FTian:2016} proposed to use RFID and blockchain technology to establish a supply chain platform from agriculture to food production in China. This system has demonstrated to guarantee the traceability of food supply-chain data. Meanwhile, the work of \cite{Sander:BFJ2018} shows that blockchain technologies can help to improve food safety via the provision of the traceable food products.  
Moreover, it is shown in \cite{Rafael:ICCSA18} that the integration of blockchain in food supply chain can allow customers to track the whole process of food production. Authors also gave a user case of using blockchain for the organic coffee industry in Colombian. Furthermore, \cite{QLin:Access19} proposes a food safety traceability system based on the blockchain and Electronic Product Code (EPC) IoT tags. In particular, this system can prevent data tampering and privacy exposure via smart contracts. A prototype of the proposed architecture has been implemented to demonstrate the effectiveness. 

\subsection{Smart grid}

The appearance of distributed renewable energy resources is reshaping the role of energy consumers from pure consumers to \emph{prosumers} who can also generate energy (\eg, from renewable energy resources) in addition to consuming energy only \cite{ZHANG:AE2018}. Energy prosumers who have extra energy can sell it to other consumers. We name the energy trading between a prosumer and a consumer (\ie, peers)  as P2P energy trading. However, it is challenging to ensure the secured and trusted energy trading between two trading parties in the distributed environment. 

The appearance of blockchain technology brings the opportunities to ensure the secured P2P energy trading. Some of recent studies proposed using blockchain technologies to tackle these challenges. For example, the work in \cite{ZLi:TII2018} developed a secure energy trading system based on consortium blockchains. This system can greatly save the trading cost without going through a central broker via the distributed consensus of blockchains. Moreover, Aitzhan and Svetinovic \cite{Aitzhan:TDSC2018} developed a decentralized energy-trading system based on blockchain technology. This system demonstrated the effectiveness in protecting confidential energy-trading transaction in decentralized smart grid systems. Furthermore, the work of \cite{Claudia:Sensors18} proposed a blockchain based mechanism to provide a secure and transparent energy demand-side management on smart grid.

\subsection{Health care}

Health care becomes one of the major social-economic problems due to the aging population while it also poses new challenges in traditional healthcare services because of the limited hospital resources. The recent advances in wearable health-care devices as well as BDA in health-care data bring the opportunities in promoting the remote health-care services at home or at clinic. As a result, the burden of the hospital resources can be potentially released \cite{Wang:IEEENet16}. For example, senior citizens staying at their homes are wearing the health-care devices at their bodies. These wearable devices continuously measure and collect health-care data including heart beat rate, blood sugar and blood pressure readings. Doctors and health-care teams can access health-care data at any time and anywhere via the health-care networks. However, assessing health-care data also brings privacy and security concerns. The vulnerability of health-care devices and the heterogeneity of health-care networks pose the challenges in preserving privacy an ensuring security of health-care data.

Incorporating blockchains into health-care networks can potentially overcome the challenges in privacy preservation and security assurance of health-care data. For example, the work of \cite{Esposito:IEEECloudComp2018} shows that using blockchain technology can protect health-care data stored in cloud servers. Meanwhile, Griggs \etal \cite{Griggs2018} developed a blockchain-based system to assure the private health-care data management. In particular, the health-care data generated by medical sensors can be automatically collected and transmitted to the system via executing smart contracts, consequently supporting the real-time patient monitoring. During the whole procedure, the privacy can be preserved via underneath blockchains. Moreover, the work of \cite{Bhuiyan:BBD2018} proposed a blockchain-based solution to manage individual health-care data and support data-sharing across different hospitals, medical centers, insurance companies and patients. During the whole process, the privacy and security of health-care data can be assured. Furthermore, Sun \etal \cite{YSun:ICCCN18} put forth an attribute-based signature scheme in decentralized health-care blockchain systems. On one hand, this scheme can verify the authenticity of health-care data and identification of the health-care data owner. On the other hand, this scheme can also preserve the privacy of the health-care data owner. The recent work \cite{Rahman:Access18} presents an in-home therapy management framework integrating IoT and blockchain-based MEC scheme to provide secrecy and anonymity assurance. The experimental results on a prototype demonstrate the effectiveness of the proposed system.

In addition, blockchain also bestows the traceability to the patients who are infected by some contagious viruses such as Severe Acute Respiratory Syndrome (SARS), Middle East Respiratory Syndrome (MERS), Wuhan Novel Coronavirus~\cite{Wu:Nature2020}. In particular, the infected or suspected patients who are wearing IoT devices can be tracked in their trajectories so that countermeasures such as quarantine can be made while the privacy of patients can be protected via blockchains.

\begin{table}[t]
\centering
\caption{Comparison of applications of blockchain of things}
\label{tab:applications}
\renewcommand{\arraystretch}{1.5}
\begin{tabular}{m{2.8cm}m{5.5cm}}
\hline
\textbf{Application} & \textbf{Benefits} \\
\hline
\multirow{2}{2.8cm}{Smart manufacturing \cite{christidis2016blockchains,Kusiak:IJPR2018,JWan:TII19}} & \checkmark Improving interoperability \\
&  \checkmark  Automating P2P business trading \\
& \checkmark  Reducing cost for trusted third party  \\
\hline

\multirow{3}{2.8cm}{Supply chain management \cite{Konstantinidis:2018,Kim:ISAFM2018,tapscott2017blockchain,Kshetri:2018,ZLi:TII19}} &  \checkmark  Assuring data provenance \\
&  \checkmark Reducing the costs in after-sale services  \\
& \checkmark Mitigating the supply chain risk  \\
\hline

\multirow{2}{2.8cm}{Food industry \cite{DTse:IEEM2017,FTian:2016,Sander:BFJ2018,Rafael:ICCSA18,QLin:Access19}} & \checkmark Improving data traceability \\
& \checkmark Enhancing food safety \\
\hline

\multirow{3}{2.8cm}{Smart grid \cite{ZHANG:AE2018,ZLi:TII2018,Aitzhan:TDSC2018,Claudia:Sensors18}} &\checkmark Securing energy trading\\
& \checkmark Improving transparency \\
& \checkmark Preserving privacy\\
\hline

\multirow{3}{2.8cm}{Health care \cite{Wang:IEEENet16,Esposito:IEEECloudComp2018,Griggs2018,Bhuiyan:BBD2018,YSun:ICCCN18,Rahman:Access18,Wu:Nature2020}} & \checkmark Assuring security \\
& \checkmark  Preserving privacy   \\
& \checkmark  Verifying authenticity   \\
\hline

\multirow{3}{2.8cm}{IoV and UAVs \cite{ZYang:IOTJ2018,HLiu:IEEENet2018,JKang:TII2017,JKang:IoTJ19,YDai:WCMag2019,YZeng:ComMag2016,kimchi2017unmanned,WANG:AC2016,Cheng:ComMag2018,Kapitonov:REDUAS2017,kumar2018unmanned}}& \checkmark Assuring trustworthiness of messages \\
& \checkmark Securing energy-trading in electric vehicles \\
& \checkmark Guaranteeing mutual-confidence among UAVs  \\
\hline
\end{tabular}
\end{table}%

\subsection{Internet of vehicles and unmanned aerial vehicles}

Internet of vehicles (IoV) essentially integrates vehicle-to-vehicle networks, vehicle-to-roadside networks, vehicle-to-infrastructure networks and vehicle-to-pedestrian networks. The decentralization, heterogeneity and non-trustworthiness of IoV pose the challenges in securing message-transmission and transaction-execution. Integrating blockchain with IoV can tackle the above challenges. For example, the work of \cite{ZYang:IOTJ2018} developed a trust-management platform in IoV on top of blockchains. In particular, the trustworthiness of messages can be validated via PoW/PoS consensus executed by Roadside Units (RSUs). Moreover, blockchain tehcnologies can be used to protect both the energy and information interactions between electric vehicles \cite{HLiu:IEEENet2018} and hybrid electric vehicles in smart grids \cite{JKang:TII2017,JKang:IoTJ19}. In the future, incorporating artificial intelligence, mobile edge computing and blockchain can further optimize the resource allocation in IoVs \cite{YDai:WCMag2019}.

Recently, unmanned aerial vehicles (UAVs) communication networks can compensate in-sufficient coverage of wireless communication networks \cite{YZeng:ComMag2016}. Meanwhile, UAVs can also be used to deliver product items \cite{kimchi2017unmanned} and acquire real-time traffic flow data \cite{WANG:AC2016}. Moreover, the recent study of\cite{Cheng:ComMag2018} also shows that UAVs can be used to support content-centric networking and mobile edge computing. However, it is challenging to assure the trustworthiness in decentralized non-trusted UAV-networks and restrict the misbehaving UAVs \cite{BLi:IoTJ19}. The integration of blockchain technology with UAV-networks can guarantee the mutual-confidence among UAVs. The work of \cite{Kapitonov:REDUAS2017} developed an autonomous platform based on Ethereum blockchain to provide the trust-management of UAVs. Moreover, IBM \cite{kumar2018unmanned} recently applied for a patent to develop a blockchain-based system to preserve privacy and assure security of UAV data. In particular, blocks in blockchains will store the information related to UAVs including model type, manufacturer, proximity to restricted region. Consequently, the misbehavior of UAVs can be detected and identified in time.

\emph{Summary.}
Table \ref{tab:applications} summarizes major BCoT applications. In particular, it is shown in Table \ref{tab:applications} that incorporating blockchain with IoT can bring a number of benefits in the aforementioned applications. In summary, BCoT has merits like reducing the cost for trusted third party, assuring security, improving data traceability, verifying the data authenticity and preserving privacy.

\section{Open research issues of Blockchain of Things}
\label{sec:chall-ibot}

\begin{figure*}[t]
\centering
\includegraphics[width=16.8cm]{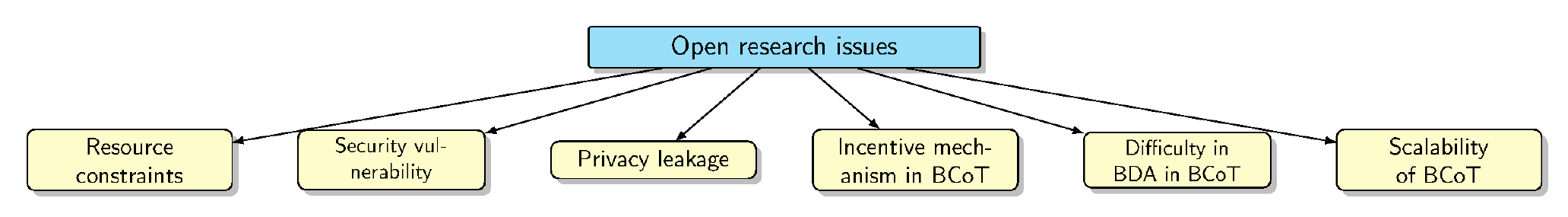}
\caption{Open research issues for blockchain of things}
\label{fig:futuredir}
\end{figure*}

Although the convergence of blockchain and IoT brings a number opportunities in upgrading the industry, there are many challenges to be addressed before the potentials of BCoT can be fully unleashed. In this section, we identify several major challenges in incorporating blockchain into IoT and discuss the potential solutions. Fig. \ref{fig:futuredir} summarizes the open research issues for blockchain of things.

\subsection{Resource constraints}

Most of IoT devices are resource-constrained. For example, sensors, RFID tags and smart meters have inferior computing capability, limited storage space, low battery power and poor network connection capability. However, the decentralized consensus algorithms of blockchains often require extensive computing power and energy consumption. For example, PoW in Bitcoin is shown to have high energy consumption \cite{REYNA:2018}. Therefore, the consensus mechanisms with  huge energy consumption may not be feasible to low-power IoT devices. 

On the other hand, the bulky size of blockchain data also results in infeasibility of fully deploying blockchains across IoT. For example, the Bitcoin blockchain size almost reaches 185 GB by the end of September 2018. It is impossible to fully store the whole blockchain at each IoT device. Meanwhile, the massive IoT data generated in nearly real time manner makes this status quo even worse. Moreover, blockchains are mainly designed for a scenario with the stable network connection, which may not be feasible for IoT that often suffers from the poor network connection of IoT devices and the unstable network due to the failure of nodes (\eg, battery depletion).

\emph{Potential solutions.} Incorporating MEC and cloud computing technologies into BCoT may potentially overcome resource constraints of IoT devices. For example, cloud servers or some MEC servers may serve as full nodes that store the whole blockchain data and participate in most of blockchain operations, such as initiating transactions, validating transactions (\ie, mining) while IoT devices may serve as lightweight nodes that only store partial blockchain data (even hash value of blockchain data) and undertake some less-computational-intensive tasks (such as initiating transactions) \cite{YDai:TVT18}. The orchestration of MEC and cloud computing becomes an important issue in the sense of allocating resource in BCoT \cite{Tran:ComMag17}.

\subsection{Security vulnerability}

Although incorporating blockchain technologies into IoT can improve the security of IoT via the encryption and digital signature brought by blockchains, the security is still a major concern for BCoT due to the vulnerabilities of IoT systems and blockchain systems. 

On one hand, there is a growing trend in deploying wireless networks into industrial environment due to the feasibility and scalability of wireless communication systems. However, the open wireless medium also makes IoT suffering from the security breaches such as passive eavesdropping \cite{li2016analytical}, jamming, replaying attacks \cite{JLin:2017}. Moreover, due to the resource constraints of IoT devices, conventional heavy-weighted encryption algorithms may not be feasible to IoT \cite{YYang:IoTJ2017}. In addition, it is also challenging to manage the keys (which are crucial to encryption algorithms) in distributed environment. 

Meanwhile, blockchain systems also have their own security vulnerabilities such as program defects of smart contracts \cite{LI:FGCS2017}. In particular, it is shown in \cite{apostolaki2017hijacking} that the malicious users can exploit Border Gateway Protocol (BGP) routing scheme to hijack blockchain messages, thereby resulting in the higher delay of block broadcasting. The work of \cite{ADHAMI2018} also shows that a Decentralized Autonomous Organization (DAO) attack  stole \$50 million worth of Ethereum by leveraging the vulnerability of smart contracts.

\emph{Potential solutions.} Security vulnerabilities of BCoT can be remedied via either the security enhancement of IoT systems or loophole repairing of blockchain. For example, cooperative jamming scheme \cite{LHu:IoTJ18} was explored to improve the security of IoT systems while no extra hardware is required for existing IoT nodes. Meanwhile,  \cite{WHu:IoTJ19} exploits key generations based on reciprocity and randomness of wireless channels in Long Range (LoRa) IoT network. In the perspective of repairing blockchain loopholes, there are also some advances. In particular, the recent work of \cite{Apostolaki:2018} proposes a secure relaying-network for blockchains, namely SABRE, which can prevent blockchain from BGP routing attacks. Regarding DAO attacks, Corda and Stellar trade the expressiveness for the verifiability of smart contracts \cite{Dinh:TKDE2018} so as to avoid DAO attacks.

\subsection{Privacy leakage}

Blockchain technologies have some mechanisms to preserve a certain data privacy of transaction records saved in blockchains. For example, transactions are made in Bitcoin via IP addresses instead of users' real identities thereby ensuring a certain anonymity. Moreover, one-time accounts are generated in Bitcoin to achieve the anonymity of users. However, these protection schemes are not robust enough. For example, it is shown in \cite{MConti:CST2018} that user pseudonyms can be cracked via learning and inferring the multiple transactions associated with one common user. In addition, the full storage of transaction data on blockchain can also lead to the potential privacy leakage as indicated in \cite{DORRI:FGCS2019}.

\emph{Potential solutions.} Recently, mixed coins are proposed to confuse attackers so that they cannot infer the exact number of real coins spent by a transaction. However, recent study \cite{moser2018empirical} demonstrates the weakness of the coin-mixed schemes via extensive realistic experiments based on Monero\footnote{A private digital currency platform (https://getmonero.org/)}. Moreover, the actual transaction can be deduced by leveraging the vulnerability of the coin-mixed schemes. The work of \cite{DORRI:FGCS2019} presents a memory optimized and flexible blockchain data storage scheme, which can somewhat reduce the privacy leakage risk.

\subsection{Incentive mechanism in BCoT}


An appropriate incentive mechanism is a benign stimulus to blockchain systems. For example, a number of Bitcoins (BTC) will be rewarded to a miner who first solves the computationally-difficult task. Meanwhile, a transaction in Ethereum will be charged with a given fee (\ie, gas) to pay the miners for the execution of contracts. Therefore, there are two issues in designing incentive mechanisms in blockchains: 1) the reward for proving (or mining) a block and 2) the compensation for processing a transaction (or a contract). 

However, it is challenging to design a proper incentive mechanism for BCoT to fulfill the requirements of different applications. Take digital currency platforms as an example, where miners are keen on the price of digital currency. For instance, the BTC reward for a generated block will be halved every 210,000 blocks \cite{saito2018make}. The reward decrement will discourage miners to contribute to the solution of the puzzle consequently migrating to other blockchain platforms. How to design a proper rewarding and publishing mechanism of digital currency is necessary to ensure the stability of blockchain systems. 

\emph{Potential solutions.} On the other hand, the reputation and honesty is an impetus to users in private or consortium blockchain systems. Therefore, going beyond digital currency, reputation credits can be used as incentives in the scenarios like personal reputation systems\cite{yasin2016online}, sharing economy \cite{bogner2016decentralised},  data provenance \cite{Liang:CCGRID2017} and the medication supply chain \cite{Darryl:ACT2017}. The recent work \cite{huang2019repchain} presents RepChain, which exploits the reputation of each node to develop the incentive mechanism.

\subsection{Difficulty in BDA in BCoT}

There is a surge of big volume of IoT data generated in nearly real time fashion. The IoT data exhibits in massive volume, heterogeneity and huge business value. Big data analytics on IoT data can extract hidden values and make intelligent decisions. However, it is challenging for apply conventional big data analytics schemes in BCoT due to the following reasons:
\begin{itemize}
\item \emph{Conventional BDA schemes cannot be applied to IoT devices due to the resource limitations}. Since IoT devices have inferior computing capability, the complicated BDA schemes cannot be deployed at IoT devices directly.  Moreover, the bulky size of blockchain data also leads to the infeasibility of the local storage of blockchain data at IoT devices. Although cloud computing can address these issues, uploading the data to remote cloud servers can also result in the privacy-breach and the long-latency \cite{wang2015cloud}. 

\item \emph{It is difficult to conduct data analytics on anonymous blockchain data.} Blockchain technologies can protect data privacy via encryption and digital signature on data records. However, it often requires the data decryption before conducting data analytics. Nevertheless, the decryption process is often time-consuming thereby resulting in the  inefficiency of data analytics \cite{xkxiao:icde2018}. It is challenging to design data analytics schemes on blockchain data without decryption.

\end{itemize}

\emph{Potential solutions.} MEC is serving as a crucial complement to cloud computing by offloading computing tasks from distant cloud servers to MEC in approximation to users. As a result, MEC can improve the response, privacy-preservation and context-awareness in contrast to cloud computing. Therefore, offloading BDA tasks to MEC servers can potentially solve the privacy-leakage and long latency issue of cloud computing with blockchain \cite{Dai:2019}. Regarding data analytics on anonymous blockchain data, there are some recent advances: 1) complex network-based community detection \cite{Remy:2018} to identify multiple addresses associated with an identical user, 2) feature extraction of transaction patterns of Bitcoin blockchain data to identify payment relationships \cite{Tasca:JRF18}, 3) analysis of user accounts and operation codes on Ethereum to detect Ponzi fraud behavior \cite{Chen:2018}.

\subsection{Scalability of BCoT}

The scalability of incumbent blockchains also limits the wide usage of blockchains in large scale IoT. The scalability of blockchains can be measured by the \emph{throughput} of transactions per second against the number of IoT nodes and the number of concurrent workloads \cite{Dinh:SIGMOD2017,Dinh:TKDE2018}. Many blockchain systems are suffering from the poor throughput. For example, it is shown in \cite{croman2016scaling} that Bitcoin can only process seven transactions per second. In contrast, VISA can process nearly 2,000 transactions per second and PayPal has the throughput of 170 transactions per second \cite{Vermeulen:2017,albrecht2018dynamics}. Ref. \cite{Conoscenti:AICCSA16} shows that Bitcoin blockchain may not be suitable for IoT due to the poor scalability. In summary, the incumbent blockchain systems may not be suitable for the applications with a large volume of transactions especially for IoT. 

\emph{Potential solutions.} There are two possible directions in improving the scalability of blockchains in IoT: 1) designing more scalable consensus algorithms and 2) constructing private or consortium blockchains for IoT. Regarding 1), we can choose the  consensus-localization strategy to improve the throughput of transactions. Meanwhile, we may implement some new blockchain structures such as directed acyclic graph (DAG) \cite{lewenberg2015inclusive} to allow the non-conflicting blocks from the side-chain to be assembled with the main chain, consequently reducing the cost for resolving bifurcation. In addition, we may consider integrating PoW with PBFT to improve the throughput of PoW similar to Sharding Protocol proposed in \cite{Luu:2016}, in which less computational-extensive puzzle is first solved in PoW and consensus is then reached in multiple small groups. 

Regarding 2), transactions in private and consortium blockchains can be processed much faster than public blockchains due to the fully-controlled systems and the limited number of permitted users. Meanwhile, the consensus can also be easily reached in private and consortium blockchains. Moreover, the fully-controlled blockchains also fulfill the requirement that an enterprise needs to have a control on different strategic sectors, \eg, ERP, MES, PLM and CRM systems \cite{Esposito:IEEECloudComp2016,Dinh:TKDE2018}. Though there are some attempts such as GemOS \cite{GemOS}, Multichain \cite{MultiChain} and Hyperledger \cite{hybperledge:2015}, more mature private and consortium blockchain platforms serving for specific industrial sectors are still expected in the future.

\section{Conclusion}
\label{sec:conc}


The incumbent Internet of Things (IoT) systems are facing a number of challenges including heterogeneity, poor interoperability, resource constraints, privacy and security vulnerability. The recent appearance of blockchain technologies essentially offers a solution to the issues with the enhanced interoperability, privacy, security, traceability and reliability. 

In this paper, we investigate integrating blockchain with IoT. We name such synthesis of blockchain and IoT as BCoT. We provide a comprehensive survey on BCoT. In particular, we first briefly introduce internet of things and blockchain technology. We then discuss the opportunities of BCoT and depict the architecture of BCoT. We next outline the research issues in blockchain for next-generation networks. We further discuss the potential applications of BCoT and outline the open research directions in BCoT.



%
\section*{Acknowledgement}
This work was supported by the National Key Research and Development Program (2016YFB1000101), the National Natural Science Foundation of China (61722214 and U1811462), Macao Science and Technology Development Fund under Grant No. 0026/2018/A1, and the Program for Guangdong Introducing Innovative and Entrepreneurial Teams (2016ZT06D211). In addition, this project has also received funding from the European Union's Horizon 2020 research and innovation programme under the Marie Sk\l{}odowska-Curie grant agreement No 824019. The authors would like to thank Gordon K.-T. Hon for his constructive comments.

\bibliography{IEEEabrv,reference}

\begin{thebibliography}{100}
\providecommand{\url}[1]{#1}
\csname url@samestyle\endcsname
\providecommand{\newblock}{\relax}
\providecommand{\bibinfo}[2]{#2}
\providecommand{\BIBentrySTDinterwordspacing}{\spaceskip=0pt\relax}
\providecommand{\BIBentryALTinterwordstretchfactor}{4}
\providecommand{\BIBentryALTinterwordspacing}{\spaceskip=\fontdimen2\font plus
\BIBentryALTinterwordstretchfactor\fontdimen3\font minus
  \fontdimen4\font\relax}
\providecommand{\BIBforeignlanguage}[2]{{%
\expandafter\ifx\csname l@#1\endcsname\relax
\typeout{** WARNING: IEEEtran.bst: No hyphenation pattern has been}%
\typeout{** loaded for the language `#1'. Using the pattern for}%
\typeout{** the default language instead.}%
\else
\language=\csname l@#1\endcsname
\fi
#2}}
\providecommand{\BIBdecl}{\relax}
\BIBdecl

\bibitem{Lade:IEEEIS2018}
P.~Lade, R.~Ghosh, and S.~Srinivasan, ``{Manufacturing Analytics and Industrial
  Internet of Things},'' \emph{IEEE Intelligent Systems}, vol.~32, no.~3, pp.
  74--79, May 2017.

\bibitem{Dorri:PercomWorkshp2017}
A.~Dorri, S.~S. Kanhere, R.~Jurdak, and P.~Gauravaram, ``{Blockchain for IoT
  security and privacy: The case study of a smart home},'' in \emph{2017 IEEE
  International Conference on Pervasive Computing and Communications Workshops
  (PerCom Workshops)}, March 2017, pp. 618--623.

\bibitem{Zhang:P2PNA2017}
\BIBentryALTinterwordspacing
Y.~Zhang and J.~Wen, ``{The IoT electric business model: Using blockchain
  technology for the internet of things},'' \emph{Peer-to-Peer Networking and
  Applications}, vol.~10, no.~4, pp. 983--994, Jul 2017. [Online]. Available:
  \url{https://doi.org/10.1007/s12083-016-0456-1}
\BIBentrySTDinterwordspacing

\bibitem{Conoscenti:AICCSA16}
M.~{Conoscenti}, A.~{Vetr}\`{o}, and J.~C. {De Martin}, ``Blockchain for the
  internet of things: A systematic literature review,'' in \emph{2016 IEEE/ACS
  13th International Conference of Computer Systems and Applications (AICCSA)},
  Nov 2016, pp. 1--6.

\bibitem{BANERJEE:2017}
M.~Banerjee, J.~Lee, and K.-K.~R. Choo, ``A blockchain future for
  internet-of-things security: a position paper,'' \emph{Digital Communications
  and Networks}, vol.~4, no.~3, pp. 149 -- 160, 2018.

\bibitem{REYNA:2018}
A.~Reyna, C.~Martín, J.~Chen, E.~Soler, and M.~Díaz, ``{On blockchain and its
  integration with IoT. Challenges and opportunities},'' \emph{Future
  Generation Computer Systems}, vol.~88, pp. 173 -- 190, 2018.

\bibitem{Fernandez-Carames:Access2018}
T.~M. Fern\^{a}ndez-Caram\^{e}s and P.~Fraga-Lamas, ``A review on the use of
  blockchain for the internet of things,'' \emph{IEEE Access}, vol.~6, pp.
  32\,979--33\,001, 2018.

\bibitem{MSAli:CST2018}
\BIBentryALTinterwordspacing
M.~S. {Ali}, M.~{Vecchio}, M.~{Pincheira}, K.~{Dolui}, F.~{Antonelli}, and
  M.~H. {Rehmani}, ``Applications of blockchains in the internet of things: A
  comprehensive survey,'' \emph{IEEE Communications Surveys Tutorials},
  vol.~21, no.~2, pp. 1676 -- 1717, 2019. [Online]. Available:
  \url{https://doi.org/10.1109/COMST.2018.2886932}
\BIBentrySTDinterwordspacing

\bibitem{Panarello:Sensors18}
\BIBentryALTinterwordspacing
A.~Panarello, N.~Tapas, G.~Merlino, F.~Longo, and A.~Puliafito, ``Blockchain
  and iot integration: A systematic survey,'' \emph{Sensors}, vol.~18, no.~8,
  2018. [Online]. Available: \url{http://www.mdpi.com/1424-8220/18/8/2575}
\BIBentrySTDinterwordspacing

\bibitem{Petersen:IEEE2011}
S.~Petersen and S.~Carlsen, ``{WirelessHART Versus ISA100.11a: The Format War
  Hits the Factory Floor},'' \emph{IEEE Industrial Electronics Magazine},
  vol.~5, no.~4, pp. 23--34, Dec 2011.

\bibitem{MEKKI2018}
K.~Mekki, E.~Bajic, F.~Chaxel, and F.~Meyer, ``{A comparative study of LPWAN
  technologies for large-scale IoT deployment},'' \emph{ICT Express}, 2018.

\bibitem{MChen:IEEEAccess2017}
M.~Chen, Y.~Miao, Y.~Hao, and K.~Hwang, ``{Narrow Band Internet of Things},''
  \emph{IEEE Access}, vol.~5, pp. 20\,557--20\,577, 2017.

\bibitem{Khutsoane:IECON2017}
O.~Khutsoane, B.~Isong, and A.~M. Abu-Mahfouz, ``{IoT devices and applications
  based on LoRa/LoRaWAN},'' in \emph{IECON 2017 - 43rd Annual Conference of the
  IEEE Industrial Electronics Society}, Oct 2017, pp. 6107--6112.

\bibitem{hndai:EIS19}
H.-N. Dai, H.~Wang, G.~Xu, J.~Wan, and M.~Imran, ``Big data analytics for
  manufacturing internet of things: Opportunities, challenges and enabling
  technologies,'' \emph{Enterprise Information Systems}, 2019.

\bibitem{XLu:IEEEWC2018}
X.~Lu, D.~Niyato, H.~Jiang, D.~I. Kim, Y.~Xiao, and Z.~Han, ``{Ambient
  Backscatter Assisted Wireless Powered Communications},'' \emph{IEEE Wireless
  Communications}, vol.~25, no.~2, pp. 170--177, April 2018.

\bibitem{JZhou:ComMag2017}
J.~Zhou, Z.~Cao, X.~Dong, and A.~V. Vasilakos, ``{Security and Privacy for
  Cloud-Based IoT: Challenges},'' \emph{IEEE Communications Magazine}, vol.~55,
  no.~1, pp. 26--33, January 2017.

\bibitem{Roman:CN2013}
R.~Roman, J.~Zhou, and J.~Lopez, ``On the features and challenges of security
  and privacy in distributed internet of things,'' \emph{Comput. Netw.},
  vol.~57, no.~10, pp. 2266--2279, Jul. 2013.

\bibitem{JHe:IoTJ18}
J.~{He}, J.~{Wei}, K.~{Chen}, Z.~{Tang}, Y.~{Zhou}, and Y.~{Zhang}, ``Multitier
  fog computing with large-scale iot data analytics for smart cities,''
  \emph{IEEE Internet of Things Journal}, vol.~5, no.~2, pp. 677--686, April
  2018.

\bibitem{zibin2016blockchain}
Z.~Zheng, S.~Xie, H.-N. Dai, X.~Chen, and H.~Wang, ``Blockchain challenges and
  opportunities: A survey,'' \emph{International Journal of Web and Grid
  Services}, vol.~14, no.~4, pp. 352 -- 375, 2018.

\bibitem{castro1999practical}
C.~Miguel and L.~Barbara, ``{Practical Byzantine fault tolerance},'' in
  \emph{Proceedings of the Third Symposium on Operating Systems Design and
  Implementation}, vol.~99, New Orleans, USA, 1999, pp. 173--186.

\bibitem{LI:FGCS2017}
X.~Li, P.~Jiang, T.~Chen, X.~Luo, and Q.~Wen, ``A survey on the security of
  blockchain systems,'' \emph{Future Generation Computer Systems}, 2017.

\bibitem{MConti:CST2018}
M.~Conti, S.~K. E, C.~Lal, and S.~Ruj, ``{A Survey on Security and Privacy
  Issues of Bitcoin},'' \emph{IEEE Communications Surveys Tutorials}, pp. 1--1,
  2018.

\bibitem{chase2018analysis}
B.~Chase and E.~MacBrough, ``{Analysis of the XRP Ledger consensus protocol},''
  \emph{arXiv preprint arXiv:1802.07242}, 2018.

\bibitem{Gilad:SOSP2017}
Y.~Gilad, R.~Hemo, S.~Micali, G.~Vlachos, and N.~Zeldovich, ``Algorand: Scaling
  byzantine agreements for cryptocurrencies,'' in \emph{Proceedings of the 26th
  Symposium on Operating Systems Principles}.\hskip 1em plus 0.5em minus
  0.4em\relax ACM, 2017, pp. 51--68.

\bibitem{FRYu:Access18}
F.~R. Yu, J.~Liu, Y.~He, P.~Si, and Y.~Zhang, ``{Virtualization for Distributed
  Ledger Technology (vDLT)},'' \emph{IEEE Access}, vol.~6, pp.
  25\,019--25\,028, 2018.

\bibitem{Dinh:SIGMOD2017}
\BIBentryALTinterwordspacing
T.~T.~A. Dinh, J.~Wang, G.~Chen, R.~Liu, B.~C. Ooi, and K.-L. Tan,
  ``Blockbench: A framework for analyzing private blockchains,'' in
  \emph{Proceedings of the 2017 ACM International Conference on Management of
  Data}, ser. SIGMOD '17.\hskip 1em plus 0.5em minus 0.4em\relax New York, NY,
  USA: ACM, 2017, pp. 1085--1100. [Online]. Available:
  \url{http://doi.acm.org/10.1145/3035918.3064033}
\BIBentrySTDinterwordspacing

\bibitem{Zyskind:IEEESPW15}
G.~{Zyskind}, O.~{Nathan}, and A.~{Pentland}, ``{Decentralizing Privacy: Using
  Blockchain to Protect Personal Data},'' in \emph{2015 IEEE Security and
  Privacy Workshops}, May 2015, pp. 180--184.

\bibitem{Chawathe2019}
S.~S. Chawathe, \emph{Clustering Blockchain Data}.\hskip 1em plus 0.5em minus
  0.4em\relax Cham: Springer International Publishing, 2019, pp. 43--72.

\bibitem{ream2016upgrading}
\BIBentryALTinterwordspacing
J.~Ream, Y.~Chu, and D.~Schatsky, ``Upgrading blockchains: Smart contract use
  cases in industry,'' Deloitte Press, 2016. [Online]. Available:
  \url{https://www2.deloitte.com/insights/us/en/focus/signals-for-strategists/using-blockchain-for-smart-contracts.html}
\BIBentrySTDinterwordspacing

\bibitem{szabo1997idea}
\BIBentryALTinterwordspacing
N.~Szabo, ``The idea of smart contracts,'' \emph{Nick Szabo's Papers and
  Concise Tutorials}, 1997. [Online]. Available:
  \url{http://www.fon.hum.uva.nl/rob/Courses/InformationInSpeech/CDROM/Literature/LOTwinterschool2006/szabo.best.vwh.net/smart_contracts_2.html}
\BIBentrySTDinterwordspacing

\bibitem{ZZheng:FGCS20}
\BIBentryALTinterwordspacing
Z.~Zheng, S.~Xie, H.-N. Dai, W.~Chen, X.~Chen, J.~Weng, and M.~Imran, ``An
  overview on smart contracts: Challenges, advances and platforms,''
  \emph{Future Generation Computer Systems}, vol. 105, pp. 475--491, December
  2020. [Online]. Available: \url{https://doi.org/10.1016/j.future.2019.12.019}
\BIBentrySTDinterwordspacing

\bibitem{Idelberger:2016}
F.~Idelberger, G.~Governatori, R.~Riveret, and G.~Sartor, ``Evaluation of
  logic-based smart contracts for blockchain systems,'' in \emph{International
  Symposium on Rules and Rule Markup Languages for the Semantic Web
  (RuleML)}.\hskip 1em plus 0.5em minus 0.4em\relax Springer, 2016, pp.
  167--183.

\bibitem{Sillaber2017}
C.~Sillaber and B.~Waltl, ``Life cycle of smart contracts in blockchain
  ecosystems,'' \emph{Datenschutz und Datensicherheit - DuD}, vol.~41, no.~8,
  pp. 497--500, Aug 2017.

\bibitem{koulu2016blockchains}
R.~Koulu, ``Blockchains and online dispute resolution: smart contracts as an
  alternative to enforcement,'' \emph{SCRIPTed}, vol.~13, p.~40, 2016.

\bibitem{nakamoto2008bitcoin}
\BIBentryALTinterwordspacing
S.~Nakamoto, ``Bitcoin: A peer-to-peer electronic cash system,'' 2008.
  [Online]. Available: \url{https://bitcoin.org/bitcoin.pdf}
\BIBentrySTDinterwordspacing

\bibitem{Ethereum}
\BIBentryALTinterwordspacing
``{Ethereum: Blockchain APP Platform}s.'' [Online]. Available:
  \url{https://www.ethereum.org/}
\BIBentrySTDinterwordspacing

\bibitem{GemOS}
\BIBentryALTinterwordspacing
``{GemOS}: the blockchain operating system.'' [Online]. Available:
  \url{https://enterprise.gem.co/}
\BIBentrySTDinterwordspacing

\bibitem{MultiChain}
\BIBentryALTinterwordspacing
``{MultiChain}: Open platform for building blockchains.'' [Online]. Available:
  \url{https://www.multichain.com/}
\BIBentrySTDinterwordspacing

\bibitem{hybperledge:2015}
\BIBentryALTinterwordspacing
``Hyperledger project,'' 2015. [Online]. Available:
  \url{https://www.hyperledger.org/}
\BIBentrySTDinterwordspacing

\bibitem{Xu:ICSA2017}
X.~Xu, I.~Weber, M.~Staples, L.~Zhu, J.~Bosch, L.~Bass, C.~Pautasso, and
  P.~Rimba, ``A taxonomy of blockchain-based systems for architecture design,''
  in \emph{IEEE International Conference on Software Architecture (ICSA)},
  2017, pp. 243--252.

\bibitem{consortium}
\BIBentryALTinterwordspacing
``Consortium chain development.'' [Online]. Available:
  \url{https://github.com/ethereum/wiki/wiki/Consortium-Chain-Development}
\BIBentrySTDinterwordspacing

\bibitem{Johnson:2001}
D.~Johnson, A.~Menezes, and S.~Vanstone, ``{The Elliptic Curve Digital
  Signature Algorithm (ECDSA)},'' \emph{International Journal of Information
  Security}, vol.~1, no.~1, pp. 36--63, 2001.

\bibitem{christidis2016blockchains}
K.~Christidis and M.~Devetsikiotis, ``Blockchains and smart contracts for the
  internet of things,'' \emph{IEEE Access}, vol.~4, pp. 2292--2303, 2016.

\bibitem{QLu:IEEESoftware2017}
Q.~Lu and X.~Xu, ``Adaptable blockchain-based systems: A case study for product
  traceability,'' \emph{IEEE Software}, vol.~34, no.~6, pp. 21--27, 2017.

\bibitem{zhang2015iot}
Y.~Zhang and J.~Wen, ``{An IoT electric business model based on the protocol of
  bitcoin},'' in \emph{Proceedings of 18th International Conference on
  Intelligence in Next Generation Networks (ICIN)}, 2015, pp. 184--191.

\bibitem{MASSARO:2017}
\BIBentryALTinterwordspacing
M.~Massaro, ``Next generation of radio spectrum management: Licensed shared
  access for 5g,'' \emph{Telecommunications Policy}, vol.~41, no.~5, pp. 422 --
  433, 2017, optimising Spectrum Use. [Online]. Available:
  \url{http://www.sciencedirect.com/science/article/pii/S0308596117301416}
\BIBentrySTDinterwordspacing

\bibitem{FCC:whitepaper18}
J.~Eggerton, ``{FCC's Rosenworcel Talks Up 6G},''
  \url{https://www.multichannel.com/news/fccs-rosenworcel-talks-up-6g}, Tech.
  Rep., September 2018.

\bibitem{Saracco:6G18}
R.~Saracco, ``{Let's start talking about 6G!}''
  \url{http://sites.ieee.org/futuredirections/2018/01/25/lets-start-talking-about-6g/},
  Tech. Rep., January 2018.

\bibitem{Gatherer:6G18}
A.~Gatherer, ``{What Will 6G Be?}''
  \url{https://www.comsoc.org/publications/ctn/what-will-6g-be}, Tech. Rep.,
  June 2018.

\bibitem{Seppo:2018}
S.~Yrj{\"o}l{\"a}, ``Analysis of blockchain use cases in the citizens broadband
  radio service spectrum sharing concept,'' in \emph{Cognitive Radio Oriented
  Wireless Networks}.\hskip 1em plus 0.5em minus 0.4em\relax Cham: Springer
  International Publishing, 2018, pp. 128--139.

\bibitem{Kotobi:VTM2108}
K.~Kotobi and S.~G. Bilen, ``{Secure Blockchains for Dynamic Spectrum Access: A
  Decentralized Database in Moving Cognitive Radio Networks Enhances Security
  and User Access},'' \emph{{IEEE Vehicular Technology Magazine}}, vol.~13,
  no.~1, pp. 32--39, March 2018.

\bibitem{Kure:IoTJ19}
\BIBentryALTinterwordspacing
E.~H.~H. {Kure}, P.~{Engelstad}, S.~{Maharjan}, S.~{Gjessing}, and Y.~{Zhang},
  ``Distributed uplink offloading for iot in 5g heterogeneous networks under
  private information constraints,'' \emph{IEEE Internet of Things Journal},
  vol.~6, no.~4, pp. 6151 -- 6164, 2019. [Online]. Available:
  \url{https://doi.org/10.1109/JIOT.2018.2886703}
\BIBentrySTDinterwordspacing

\bibitem{Huawei:whitepaper17}
E.~Langberg, ``{Blockchains in Mobile Networks},''
  \url{https://e.huawei.com/hk/publications/global/ict$\_$insights/201703141505/},
  Tech. Rep.~21, March 2017.

\bibitem{Sheng:ICBC2018}
S.~He, C.~Xing, and L.-J. Zhang, ``A business-oriented schema for blockchain
  network operation,'' in \emph{Blockchain -- ICBC 2018}, S.~Chen, H.~Wang, and
  L.-J. Zhang, Eds.\hskip 1em plus 0.5em minus 0.4em\relax Cham: Springer
  International Publishing, 2018, pp. 277--284.

\bibitem{hndai:BDAWireless2019}
\BIBentryALTinterwordspacing
H.-N. Dai, R.~C.-W. Wong, H.~Wang, Z.~Zheng, and A.~V. Vasilakos, ``Big data
  analytics for large scale wireless networks: Challenges and opportunities,''
  \emph{ACM Computing Surveys}, vol.~52, no.~5, 2019. [Online]. Available:
  \url{https://doi.org/10.1145/3337065}
\BIBentrySTDinterwordspacing

\bibitem{Bera:IoTJ2017}
S.~Bera, S.~Misra, and A.~V. Vasilakos, ``Software-defined networking for
  internet of things: A survey,'' \emph{IEEE Internet of Things Journal},
  vol.~4, no.~6, pp. 1994--2008, Dec 2017.

\bibitem{Kalkan:ComMag2017}
K.~Kalkan and S.~Zeadally, ``Securing internet of things with software defined
  networking,'' \emph{IEEE Communications Magazine}, vol.~56, no.~9, pp.
  186--192, September 2018.

\bibitem{Sharma:ComMag2017}
P.~K. Sharma, S.~Singh, Y.~Jeong, and J.~H. Park, ``Distblocknet: A distributed
  blockchains-based secure sdn architecture for iot networks,'' \emph{IEEE
  Communications Magazine}, vol.~55, no.~9, pp. 78--85, 2017.

\bibitem{Alvarenga:NOMS18}
I.~D. {Alvarenga}, G.~A.~F. {Rebello}, and O.~C. M.~B. {Duarte}, ``Securing
  configuration management and migration of virtual network functions using
  blockchain,'' in \emph{NOMS 2018 - 2018 IEEE/IFIP Network Operations and
  Management Symposium}, April 2018, pp. 1--9.

\bibitem{Afolabi:CST2018}
I.~Afolabi, T.~Taleb, K.~Samdanis, A.~Ksentini, and H.~Flinck, ``{Network
  Slicing and Softwarization: A Survey on Principles, Enabling Technologies,
  and Solutions},'' \emph{IEEE Communications Surveys Tutorials}, vol.~20,
  no.~3, pp. 2429--2453, 2018.

\bibitem{Esposito:IEEECloudComp2016}
C.~Esposito, A.~Castiglione, B.~Martini, and K.~K.~R. Choo, ``Cloud
  manufacturing: Security, privacy, and forensic concerns,'' \emph{IEEE Cloud
  Computing}, vol.~3, no.~4, pp. 16--22, July 2016.

\bibitem{Ortega:VTM2018}
V.~Ortega, F.~Bouchmal, and J.~F. Monserrat, ``Trusted 5g vehicular networks:
  Blockchains and content-centric networking,'' \emph{IEEE Vehicular Technology
  Magazine}, vol.~13, no.~2, pp. 121--127, June 2018.

\bibitem{Fan:IETCom2018}
K.~Fan, Y.~Ren, Y.~Wang, H.~Li, and Y.~Yang, ``Blockchain-based efficient
  privacy preserving and data sharing scheme of content-centric network in
  5g,'' \emph{IET Communications}, vol.~12, no.~5, pp. 527--532, 2018.

\bibitem{CChen:IEEENet2018}
C.~Chen, M.~Lin, and C.~Liu, ``Edge computing gateway of the industrial
  internet of things using multiple collaborative microcontrollers,''
  \emph{IEEE Network}, vol.~32, no.~1, pp. 24--32, 2018.

\bibitem{Abbas:IoTJ2018}
N.~Abbas, Y.~Zhang, A.~Taherkordi, and T.~Skeie, ``Mobile edge computing: A
  survey,'' \emph{IEEE Internet of Things Journal}, vol.~5, no.~1, pp.
  450--465, 2018.

\bibitem{ZXiong:ComMag2018}
Z.~Xiong, Y.~Zhang, D.~Niyato, P.~Wang, and Z.~Han, ``{When Mobile Blockchain
  Meets Edge Computing},'' \emph{IEEE Communications Magazine}, vol.~56, no.~8,
  pp. 33--39, August 2018.

\bibitem{MLiu:TVT18}
M.~Liu, F.~R. Yu, Y.~Teng, V.~C.~M. Leung, and M.~Song, ``Computation
  offloading and content caching in wireless blockchain networks with mobile
  edge computing,'' \emph{IEEE Transactions on Vehicular Technology}, vol.~67,
  no.~11, pp. 11\,008--11\,021, Nov 2018.

\bibitem{Yang:NetMag2017}
P.~Yang, N.~Zhang, Y.~Bi, L.~Yu, and X.~S. Shen, ``Catalyzing cloud-fog
  interoperation in 5g wireless networks: An sdn approach,'' \emph{IEEE
  Network}, vol.~31, no.~5, pp. 14--20, 2017.

\bibitem{YDai:IoTJ19}
\BIBentryALTinterwordspacing
Y.~{Dai}, D.~{Xu}, S.~{Maharjan}, and Y.~{Zhang}, ``Joint load balancing and
  offloading in vehicular edge computing and networks,'' \emph{IEEE Internet of
  Things Journal}, vol.~6, no.~3, pp. 4377 -- 4387, 2019. [Online]. Available:
  \url{https://doi.org/10.1109/JIOT.2018.2876298}
\BIBentrySTDinterwordspacing

\bibitem{ZZhou:TVT19}
\BIBentryALTinterwordspacing
Z.~{Zhou}, P.~{Liu}, J.~{Feng}, Y.~{Zhang}, S.~{Mumtaz}, and J.~{Rodriguez},
  ``Computation resource allocation and task assignment optimization in
  vehicular fog computing: A contract-matching approach,'' \emph{IEEE
  Transactions on Vehicular Technology}, vol.~68, no.~4, pp. 3113 -- 3125,
  2019. [Online]. Available: \url{https://doi.org/10.1109/TVT.2019.2894851}
\BIBentrySTDinterwordspacing

\bibitem{YYang:IoTJ18}
Y.~{Yang}, K.~{Wang}, G.~{Zhang}, X.~{Chen}, X.~{Luo}, and M.~{Zhou}, ``Meets:
  Maximal energy efficient task scheduling in homogeneous fog networks,''
  \emph{IEEE Internet of Things Journal}, vol.~5, no.~5, pp. 4076--4087, Oct
  2018.

\bibitem{Kusiak:IJPR2018}
A.~Kusiak, ``Smart manufacturing,'' \emph{International Journal of Production
  Research}, vol.~56, no. 1-2, pp. 508--517, 2018.

\bibitem{JWan:TII19}
\BIBentryALTinterwordspacing
J.~{Wan}, J.~{Li}, M.~{Imran}, D.~{Li}, and F.~{e-Amin}, ``A blockchain-based
  solution for enhancing security and privacy in smart factory,'' \emph{IEEE
  Transactions on Industrial Informatics}, vol.~15, no.~6, pp. 3652 -- 3660,
  2019. [Online]. Available: \url{https://doi.org/10.1109/TII.2019.2894573}
\BIBentrySTDinterwordspacing

\bibitem{JHuang:TII20}
J.~{Huang}, L.~{Kong}, H.-N. {Dai}, W.~{Ding}, L.~{Cheng}, G.~{Chen}, X.~{Jin},
  and P.~{Zeng}, ``Blockchain based mobile crowd sensing in industrial
  systems,'' \emph{IEEE Transactions on Industrial Informatics}, pp. 1--1,
  2020.

\bibitem{Konstantinidis:2018}
I.~Konstantinidis, G.~Siaminos, C.~Timplalexis, P.~Zervas, V.~Peristeras, and
  S.~Decker, ``Blockchain for business applications: A systematic literature
  review,'' in \emph{Business Information Systems}, W.~Abramowicz and
  A.~Paschke, Eds.\hskip 1em plus 0.5em minus 0.4em\relax Cham: Springer
  International Publishing, 2018, pp. 384--399.

\bibitem{Kim:ISAFM2018}
H.~M. Kim and M.~Laskowski, ``Toward an ontology-driven blockchain design for
  supply-chain provenance,'' \emph{Intelligent Systems in Accounting, Finance
  and Management}, vol.~25, no.~1, pp. 18--27, 2018.

\bibitem{tapscott2017blockchain}
A.~Tapscott and D.~Tapscott, ``How blockchain is changing finance,''
  \emph{Harvard Business Review}, vol.~1, 2017.

\bibitem{Kshetri:2018}
N.~Kshetri, ``1 blockchain’s roles in meeting key supply chain management
  objectives,'' \emph{International Journal of Information Management},
  vol.~39, pp. 80 -- 89, 2018.

\bibitem{ZLi:TII19}
\BIBentryALTinterwordspacing
Z.~{Li}, H.~{Guo}, W.~M. {Wang}, Y.~{Guan}, A.~{Vatankhah Barenji}, G.~Q.
  {Huang}, K.~S. {McFall}, and X.~{Chen}, ``A blockchain and automl approach
  for open and automated customer service,'' \emph{IEEE Transactions on
  Industrial Informatics}, vol.~15, no.~6, pp. 3642 -- 3651, 2019. [Online].
  Available: \url{https://doi.org/10.1109/TII.2019.2900987}
\BIBentrySTDinterwordspacing

\bibitem{DTse:IEEM2017}
D.~Tse, B.~Zhang, Y.~Yang, C.~Cheng, and H.~Mu, ``Blockchain application in
  food supply information security,'' in \emph{2017 IEEE International
  Conference on Industrial Engineering and Engineering Management (IEEM)}, Dec
  2017, pp. 1357--1361.

\bibitem{FTian:2016}
F.~Tian, ``An agri-food supply chain traceability system for china based on
  rfid amp;amp; blockchain technology,'' in \emph{13th International Conference
  on Service Systems and Service Management (ICSSSM)}, 2016, pp. 1--6.

\bibitem{Sander:BFJ2018}
F.~Sander, J.~Semeijn, and D.~Mahr, ``The acceptance of blockchain technology
  in meat traceability and transparency,'' \emph{British Food Journal}, vol.~0,
  no.~0, p. null, 2018.

\bibitem{Rafael:ICCSA18}
R.~Bett{\'i}n-D{\'i}az, A.~E. Rojas, and C.~Mej{\'i}a-Moncayo, ``Methodological
  approach to the definition of a blockchain system for the food industry
  supply chain traceability,'' in \emph{Computational Science and Its
  Applications -- ICCSA 2018}.\hskip 1em plus 0.5em minus 0.4em\relax Cham:
  Springer International Publishing, 2018, pp. 19--33.

\bibitem{QLin:Access19}
Q.~{Lin}, H.~{Wang}, X.~{Pei}, and J.~{Wang}, ``Food safety traceability system
  based on blockchain and epcis,'' \emph{IEEE Access}, vol.~7, pp.
  20\,698--20\,707, 2019.

\bibitem{ZHANG:AE2018}
\BIBentryALTinterwordspacing
C.~Zhang, J.~Wu, Y.~Zhou, M.~Cheng, and C.~Long, ``Peer-to-peer energy trading
  in a microgrid,'' \emph{Applied Energy}, vol. 220, pp. 1 -- 12, 2018.
  [Online]. Available:
  \url{http://www.sciencedirect.com/science/article/pii/S0306261918303398}
\BIBentrySTDinterwordspacing

\bibitem{ZLi:TII2018}
Z.~Li, J.~Kang, R.~Yu, D.~Ye, Q.~Deng, and Y.~Zhang, ``{Consortium Blockchain
  for Secure Energy Trading in Industrial Internet of Things},'' \emph{IEEE
  Transactions on Industrial Informatics}, vol.~14, no.~8, pp. 3690--3700, Aug
  2018.

\bibitem{Aitzhan:TDSC2018}
N.~Z. Aitzhan and D.~Svetinovic, ``Security and privacy in decentralized energy
  trading through multi-signatures, blockchain and anonymous messaging
  streams,'' \emph{IEEE Transactions on Dependable and Secure Computing},
  vol.~15, no.~5, pp. 840--852, Sept 2018.

\bibitem{Claudia:Sensors18}
C.~Pop, T.~Cioara, M.~Antal, I.~Anghel, I.~Salomie, and M.~Bertoncini,
  ``Blockchain based decentralized management of demand response programs in
  smart energy grids,'' \emph{Sensors}, vol.~18, no.~1, 2018.

\bibitem{Wang:IEEENet16}
K.~Wang, Y.~Shao, L.~Shu, C.~Zhu, and Y.~Zhang, ``Mobile big data
  fault-tolerant processing for ehealth networks,'' \emph{IEEE Network},
  vol.~30, no.~1, pp. 36--42, January 2016.

\bibitem{Esposito:IEEECloudComp2018}
C.~Esposito, A.~D. Santis, G.~Tortora, H.~Chang, and K.~R. Choo, ``Blockchain:
  A panacea for healthcare cloud-based data security and privacy?'' \emph{IEEE
  Cloud Computing}, vol.~5, no.~1, pp. 31--37, Jan 2018.

\bibitem{Griggs2018}
\BIBentryALTinterwordspacing
K.~N. Griggs, O.~Ossipova, C.~P. Kohlios, A.~N. Baccarini, E.~A. Howson, and
  T.~Hayajneh, ``Healthcare blockchain system using smart contracts for secure
  automated remote patient monitoring,'' \emph{Journal of Medical Systems},
  vol.~42, no.~7, p. 130, Jun 2018. [Online]. Available:
  \url{https://doi.org/10.1007/s10916-018-0982-x}
\BIBentrySTDinterwordspacing

\bibitem{Bhuiyan:BBD2018}
M.~Z.~A. Bhuiyan, A.~Zaman, T.~Wang, G.~Wang, H.~Tao, and M.~M. Hassan,
  ``Blockchain and big data to transform the healthcare,'' in \emph{Proceedings
  of the International Conference on Data Processing and Applications}, ser.
  ICDPA.\hskip 1em plus 0.5em minus 0.4em\relax ACM, 2018, pp. 62--68.

\bibitem{YSun:ICCCN18}
Y.~Sun, R.~Zhang, X.~Wang, K.~Gao, and L.~Liu, ``A decentralizing
  attribute-based signature for healthcare blockchain,'' in \emph{2018 27th
  International Conference on Computer Communication and Networks (ICCCN)},
  2018, pp. 1--9.

\bibitem{Rahman:Access18}
M.~A. {Rahman}, M.~S. {Hossain}, G.~{Loukas}, E.~{Hassanain}, S.~S. {Rahman},
  M.~F. {Alhamid}, and M.~{Guizani}, ``Blockchain-based mobile edge computing
  framework for secure therapy applications,'' \emph{IEEE Access}, vol.~6, pp.
  72\,469--72\,478, 2018.

\bibitem{Wu:Nature2020}
\BIBentryALTinterwordspacing
F.~Wu \emph{et~al.}, ``A new coronavirus associated with human respiratory
  disease in china,'' \emph{Nature}, 2020. [Online]. Available:
  \url{https://doi.org/10.1038/s41586-020-2008-3}
\BIBentrySTDinterwordspacing

\bibitem{ZYang:IOTJ2018}
\BIBentryALTinterwordspacing
Z.~Yang, K.~Yang, L.~Lei, K.~Zheng, and V.~C.~M. Leung, ``Blockchain-based
  decentralized trust management in vehicular networks,'' \emph{IEEE Internet
  of Things Journal}, vol.~6, no.~2, pp. 1495 -- 1505, May 2018. [Online].
  Available: \url{https://doi.org/10.1109/JIOT.2018.2836144}
\BIBentrySTDinterwordspacing

\bibitem{HLiu:IEEENet2018}
H.~Liu, Y.~Zhang, and T.~Yang, ``Blockchain-enabled security in electric
  vehicles cloud and edge computing,'' \emph{IEEE Network}, vol.~32, no.~3, pp.
  78--83, May 2018.

\bibitem{JKang:TII2017}
J.~Kang, R.~Yu, X.~Huang, S.~Maharjan, Y.~Zhang, and E.~Hossain, ``Enabling
  localized peer-to-peer electricity trading among plug-in hybrid electric
  vehicles using consortium blockchains,'' \emph{IEEE Transactions on
  Industrial Informatics}, vol.~13, no.~6, pp. 3154--3164, Dec 2017.

\bibitem{JKang:IoTJ19}
J.~{Kang}, R.~{Yu}, X.~{Huang}, M.~{Wu}, S.~{Maharjan}, S.~{Xie}, and
  Y.~{Zhang}, ``Blockchain for secure and efficient data sharing in vehicular
  edge computing and networks,'' \emph{IEEE Internet of Things Journal},
  vol.~6, no.~3, pp. 4660 -- 4670, 2019.

\bibitem{YDai:WCMag2019}
Y.~Dai, D.~Xu, S.~Maharjan, G.~Qiao, and Y.~Zhang, ``{Artificial Intelligence
  Empowered Edge Computing and Caching for Internet of Vehicles},'' \emph{IEEE
  Wireless Communications Magazine}, vol.~26, no.~3, pp. 12 -- 18, 2019.

\bibitem{YZeng:ComMag2016}
Y.~Zeng, R.~Zhang, and T.~J. Lim, ``Wireless communications with unmanned
  aerial vehicles: opportunities and challenges,'' \emph{IEEE Communications
  Magazine}, vol.~54, no.~5, pp. 36--42, May 2016.

\bibitem{kimchi2017unmanned}
G.~Kimchi, D.~Buchmueller, S.~A. Green, B.~C. Beckman, S.~Isaacs, A.~Navot,
  F.~Hensel, A.~Bar-Zeev, and S.~S. J.-M. Rault, ``Unmanned aerial vehicle
  delivery system,'' 2017, uS Patent 9,573,684.

\bibitem{WANG:AC2016}
\BIBentryALTinterwordspacing
L.~Wang, F.~Chen, and H.~Yin, ``Detecting and tracking vehicles in traffic by
  unmanned aerial vehicles,'' \emph{Automation in Construction}, vol.~72, pp.
  294 -- 308, 2016. [Online]. Available:
  \url{http://www.sciencedirect.com/science/article/pii/S0926580516300887}
\BIBentrySTDinterwordspacing

\bibitem{Cheng:ComMag2018}
N.~Cheng, W.~Xu, W.~Shi, Y.~Zhou, N.~Lu, H.~Zhou, and X.~Shen, ``Air-ground
  integrated mobile edge networks: Architecture, challenges, and
  opportunities,'' \emph{IEEE Communications Magazine}, vol.~56, no.~8, pp.
  26--32, August 2018.

\bibitem{Kapitonov:REDUAS2017}
A.~Kapitonov, S.~Lonshakov, A.~Krupenkin, and I.~Berman, ``Blockchain-based
  protocol of autonomous business activity for multi-agent systems consisting
  of uavs,'' in \emph{2017 Workshop on Research, Education and Development of
  Unmanned Aerial Systems (RED-UAS)}, 2017, pp. 84--89.

\bibitem{kumar2018unmanned}
A.~Kumar, A.~Kundu, C.~A. Pickover, and K.~Weldemariam, ``Unmanned aerial
  vehicle data management,'' 2018, uS Patent App. 15/463,147.

\bibitem{BLi:IoTJ19}
\BIBentryALTinterwordspacing
B.~{Li}, Z.~{Fei}, and Y.~{Zhang}, ``Uav communications for 5g and beyond:
  Recent advances and future trends,'' \emph{IEEE Internet of Things Journal},
  vol.~6, no.~2, pp. 2241 -- 2263, 2019. [Online]. Available:
  \url{https://doi.org/10.1109/JIOT.2018.2887086}
\BIBentrySTDinterwordspacing

\bibitem{YDai:TVT18}
Y.~{Dai}, D.~{Xu}, S.~{Maharjan}, and Y.~{Zhang}, ``{Joint Computation
  Offloading and User Association in Multi-Task Mobile Edge Computing},''
  \emph{IEEE Transactions on Vehicular Technology}, vol.~67, no.~12, pp.
  12\,313--12\,325, Dec 2018.

\bibitem{Tran:ComMag17}
T.~X. {Tran}, A.~{Hajisami}, P.~{Pandey}, and D.~{Pompili}, ``Collaborative
  mobile edge computing in 5g networks: New paradigms, scenarios, and
  challenges,'' \emph{IEEE Communications Magazine}, vol.~55, no.~4, pp.
  54--61, April 2017.

\bibitem{li2016analytical}
X.~Li, H.~Wang, H.-N. Dai, Y.~Wang, and Q.~Zhao, ``An analytical study on
  eavesdropping attacks in wireless nets of things,'' \emph{Mobile Information
  Systems}, vol. 2016, 2016.

\bibitem{JLin:2017}
J.~Lin, W.~Yu, N.~Zhang, X.~Yang, H.~Zhang, and W.~Zhao, ``A survey on internet
  of things: Architecture, enabling technologies, security and privacy, and
  applications,'' \emph{IEEE Internet of Things Journal}, vol.~4, no.~5, pp.
  1125--1142, 2017.

\bibitem{YYang:IoTJ2017}
Y.~Yang, L.~Wu, G.~Yin, L.~Li, and H.~Zhao, ``A survey on security and privacy
  issues in internet-of-things,'' \emph{IEEE Internet of Things Journal},
  vol.~4, no.~5, pp. 1250--1258, 2017.

\bibitem{apostolaki2017hijacking}
M.~Apostolaki, A.~Zohar, and L.~Vanbever, ``{Hijacking Bitcoin: Routing attacks
  on cryptocurrencies},'' in \emph{Security and Privacy (SP), IEEE Symposium
  on}.\hskip 1em plus 0.5em minus 0.4em\relax IEEE, 2017, pp. 375--392.

\bibitem{ADHAMI2018}
\BIBentryALTinterwordspacing
S.~Adhami, G.~Giudici, and S.~Martinazzi, ``Why do businesses go crypto? an
  empirical analysis of initial coin offerings,'' \emph{Journal of Economics
  and Business}, vol. 100, pp. 64 -- 75, 2018. [Online]. Available:
  \url{http://www.sciencedirect.com/science/article/pii/S0148619517302308}
\BIBentrySTDinterwordspacing

\bibitem{LHu:IoTJ18}
L.~{Hu}, H.~{Wen}, B.~{Wu}, F.~{Pan}, R.~{Liao}, H.~{Song}, J.~{Tang}, and
  X.~{Wang}, ``Cooperative jamming for physical layer security enhancement in
  internet of things,'' \emph{IEEE Internet of Things Journal}, vol.~5, no.~1,
  pp. 219--228, Feb 2018.

\bibitem{WHu:IoTJ19}
W.~{Xu}, S.~{Jha}, and W.~{Hu}, ``{LoRa-Key: Secure Key Generation System for
  LoRa-based Network},'' \emph{IEEE Internet of Things Journal}, vol.~6, no.~4,
  pp. 6404 -- 6416, 2019.

\bibitem{Apostolaki:2018}
M.~Apostolaki, G.~Marti, J.~Müller, and L.~Vanbever, ``{SABRE: Protecting
  Bitcoin against Routing Attacks},'' in \emph{Proceedings of the Network and
  Distributed System Security Symposium}, 2019, pp. 1--15.

\bibitem{Dinh:TKDE2018}
T.~T.~A. Dinh, R.~Liu, M.~Zhang, G.~Chen, B.~C. Ooi, and J.~Wang, ``Untangling
  blockchain: A data processing view of blockchain systems,'' \emph{IEEE
  Transactions on Knowledge and Data Engineering}, vol.~30, no.~7, pp.
  1366--1385, July 2018.

\bibitem{DORRI:FGCS2019}
\BIBentryALTinterwordspacing
A.~Dorri, S.~S. Kanhere, and R.~Jurdak, ``{MOF-BC: A memory optimized and
  flexible blockchain for large scale networks},'' \emph{Future Generation
  Computer Systems}, vol.~92, pp. 357 -- 373, 2019. [Online]. Available:
  \url{http://www.sciencedirect.com/science/article/pii/S0167739X17329552}
\BIBentrySTDinterwordspacing

\bibitem{moser2018empirical}
M.~M{\"o}ser, K.~Soska, E.~Heilman, K.~Lee, H.~Heffan, S.~Srivastava, K.~Hogan,
  J.~Hennessey, A.~Miller, A.~Narayanan \emph{et~al.}, ``{An Empirical Analysis
  of Traceability in the Monero Blockchain},'' \emph{Proceedings on Privacy
  Enhancing Technologies}, vol. 2018, no.~3, pp. 143--163, 2018.

\bibitem{saito2018make}
K.~Saito and M.~Iwamura, ``How to make a digital currency on a blockchain
  stable,'' \emph{arXiv preprint arXiv:1801.06771}, 2018.

\bibitem{yasin2016online}
A.~Yasin and L.~Liu, ``An online identity and smart contract management
  system,'' in \emph{Proceedings of 40th Annual Computer Software and
  Applications Conference (COMPSAC)}, vol.~2, 2016, pp. 192--198.

\bibitem{bogner2016decentralised}
A.~Bogner, M.~Chanson, and A.~Meeuw, ``A decentralised sharing app running a
  smart contract on the ethereum blockchain,'' in \emph{Proceedings of the 6th
  International Conference on the Internet of Things}, 2016, pp. 177--178.

\bibitem{Liang:CCGRID2017}
X.~Liang, S.~Shetty, D.~Tosh, C.~Kamhoua, K.~Kwiat, and L.~Njilla, ``Provchain:
  A blockchain-based data provenance architecture in cloud environment with
  enhanced privacy and availability,'' in \emph{2017 17th IEEE/ACM
  International Symposium on Cluster, Cloud and Grid Computing (CCGRID)}, 2017,
  pp. 468--477.

\bibitem{Darryl:ACT2017}
\BIBentryALTinterwordspacing
D.~G. Glover and J.~Hermans, ``\BIBforeignlanguage{English}{Improving the
  traceability of the clinical trial supply chain},''
  \emph{\BIBforeignlanguage{English}{Applied Clinical Trials}}, vol.~26,
  no.~11, pp. 36--38, November 2017. [Online]. Available:
  \url{https://search.proquest.com/docview/1984377517?accountid=28120}
\BIBentrySTDinterwordspacing

\bibitem{huang2019repchain}
C.~Huang, Z.~Wang, H.~Chen, Q.~Hu, Q.~Zhang, W.~Wang, and X.~Guan, ``Repchain:
  A reputation based secure, fast and high incentive blockchain system via
  sharding,'' 2019.

\bibitem{wang2015cloud}
P.~Wang, R.~X. Gao, and Z.~Fan, ``Cloud computing for cloud manufacturing:
  benefits and limitations,'' \emph{Journal of Manufacturing Science and
  Engineering}, vol. 137, no.~4, pp. 1--9, 2015.

\bibitem{xkxiao:icde2018}
N.~Wang, X.~Xiao, Y.~Yang, T.~D. Hoang, H.~Shin, J.~Shin, and G.~Yu,
  ``Privtrie: Effective frequent term discovery under local differential
  privacy,'' in \emph{IEEE International Conference on Data Engineering
  (ICDE)}, 2018.

\bibitem{Dai:2019}
Y.~Dai, D.~Xu, S.~Maharjan, Z.~Chen, Q.~He, and Y.~Zhang, ``Blockchain and deep
  reinforcement learning empowered intelligent 5g beyond,'' \emph{IEEE Network
  Magazine}, 2019 (in press).

\bibitem{Remy:2018}
C.~Remy, B.~Rym, and L.~Matthieu, ``Tracking bitcoin users activity using
  community detection on a network of weak signals,'' in \emph{Complex Networks
  {\&} Their Applications VI}.\hskip 1em plus 0.5em minus 0.4em\relax Cham:
  Springer International Publishing, 2018, pp. 166--177.

\bibitem{Tasca:JRF18}
P.~Tasca, A.~Hayes, and S.~Liu, ``The evolution of the bitcoin economy:
  Extracting and analyzing the network of payment relationships,'' \emph{The
  Journal of Risk Finance}, vol.~19, no.~2, pp. 94--126, 2018.

\bibitem{Chen:2018}
\BIBentryALTinterwordspacing
W.~Chen, Z.~Zheng, J.~Cui, E.~Ngai, P.~Zheng, and Y.~Zhou, ``Detecting ponzi
  schemes on ethereum: Towards healthier blockchain technology,'' in
  \emph{Proceedings of the 2018 World Wide Web Conference}, ser. WWW '18.\hskip
  1em plus 0.5em minus 0.4em\relax Republic and Canton of Geneva, Switzerland:
  International World Wide Web Conferences Steering Committee, 2018, pp.
  1409--1418. [Online]. Available:
  \url{https://doi.org/10.1145/3178876.3186046}
\BIBentrySTDinterwordspacing

\bibitem{croman2016scaling}
K.~Croman, C.~Decker, I.~Eyal, A.~E. Gencer, A.~Juels, A.~Kosba, A.~Miller,
  P.~Saxena, E.~Shi, E.~G. Sirer \emph{et~al.}, ``On scaling decentralized
  blockchains,'' in \emph{International Conference on Financial Cryptography
  and Data Security}.\hskip 1em plus 0.5em minus 0.4em\relax Springer, 2016,
  pp. 106--125.

\bibitem{Vermeulen:2017}
\BIBentryALTinterwordspacing
J.~Vermeulen, ``{Bitcoin and Ethereum vs Visa and PayPal –Transactions per
  second},'' \emph{Altcoin Today}, April 2017. [Online]. Available:
  \url{http://www.altcointoday.com/bitcoin-ethereum-vs-visa-paypal-transactions-per-second/}
\BIBentrySTDinterwordspacing

\bibitem{albrecht2018dynamics}
S.~Albrecht, S.~Reichert, J.~Schmid, J.~Str{\"u}ker, D.~Neumann, and
  G.~Fridgen, ``Dynamics of blockchain implementation-a case study from the
  energy sector,'' in \emph{Proceedings of the 51st Hawaii International
  Conference on System Sciences}, 2018.

\bibitem{lewenberg2015inclusive}
Y.~Lewenberg, Y.~Sompolinsky, and A.~Zohar, ``Inclusive block chain
  protocols,'' in \emph{International Conference on Financial Cryptography and
  Data Security}.\hskip 1em plus 0.5em minus 0.4em\relax Springer, 2015, pp.
  528--547.

\bibitem{Luu:2016}
\BIBentryALTinterwordspacing
L.~Luu, V.~Narayanan, C.~Zheng, K.~Baweja, S.~Gilbert, and P.~Saxena, ``A
  secure sharding protocol for open blockchains,'' in \emph{Proceedings of the
  2016 ACM SIGSAC Conference on Computer and Communications Security}, ser. CCS
  '16.\hskip 1em plus 0.5em minus 0.4em\relax New York, NY, USA: ACM, 2016, pp.
  17--30. [Online]. Available: \url{http://doi.acm.org/10.1145/2976749.2978389}
\BIBentrySTDinterwordspacing

\end{thebibliography}

\end{document}